\documentclass{jfm}

\usepackage{graphicx}
\usepackage{newtxtext}
\usepackage{newtxmath}
\usepackage{natbib}
\usepackage{hyperref}
\hypersetup{
    colorlinks = true,
    urlcolor   = blue,
    citecolor  = blue,
}

\newcommand{\RomanNumeralCaps}[1]
\linenumbers

\usepackage{verbatim}

\title{Small-scale dynamics and structure of free-surface turbulence}

\author{Yinghe Qi\aff{1}
  \corresp{\email{yingqi@ethz.ch}},
  Yaxing Li\aff{2}
  \corresp{\email{yaxingli@zju.edu.cn}}
 \and Filippo Coletti\aff{1}}

\affiliation{\aff{1}Department of Mechanical and Process Engineering, ETH Zurich, 8092 Zurich, Switzerland \\
\aff{2} Department of Engineering Mechanics, School of Aeronautics and Astronautics, Zhejiang University, Hangzhou, China}

\begin{document}

\maketitle

\begin{abstract}
The dynamics of small-scale structures in free-surface turbulence is crucial to large-scale phenomena in natural and industrial environments. Here we conduct experiments on the quasi-flat free surface of a zero-mean-flow turbulent water tank over the Reynolds number range \(Re_{\lambda} = 207\)--312. By seeding microscopic floating particles at high concentrations, the fine scales of the flow and the velocity gradient tensor are resolved. A kinematic relation is derived expressing the contribution of surface divergence and vorticity to the dissipation rate. The probability density functions of divergence, vorticity and strain-rate collapse once normalized by the Kolmogorov scales. Their magnitude displays strong intermittency and follows chi-square distributions with power-law tails at small values. The topology of high-intensity events and two-point statistics indicate that the surface divergence is characterized by dissipative spatial and temporal scales, while the high-vorticity and high-strain-rate regions are larger, long-lived, concurrent, and elongated. The second-order velocity structure functions obey the classic Kolmogorov scaling in the inertial range when the dissipation rate on the surface is considered, with a different numerical constant than in 3D turbulence. The cross-correlation among divergence, vorticity and strain-rate indicates that the surface-attached vortices are strengthened during downwellings and diffuse when those dissipate. Sources (sinks) in the surface velocity fields are associated with strong (weak) surface-parallel stretching and compression along perpendicular directions. The floating particles cluster over spatial and temporal scales larger than those of the sinks. These results demonstrate that, compared to 3D turbulence, in free-surface turbulence the energetic scales leave a stronger imprint on the small-scale quantities.

\end{abstract}

\begin{keywords}
Authors should not enter keywords on the manuscript, as these must be chosen by the author during the online submission process and will then be added during the typesetting process (see http://journals.cambridge.org/data/\linebreak[3]relatedlink/jfm-\linebreak[3]keywords.pdf for the full list)
\end{keywords}

\section{Introduction}\label{introduction-1}

From a cup of stirred coffee to the flow in rivers, lakes and oceans, free-surface turbulence is ubiquitous in various natural and industrial environments. The dynamics of the free surface affects the exchange of mass, momentum, and energy with the bulk, and thus plays an essential role at the global scale including the exchange of gas between the atmosphere and ocean \citep{jahne1998air,veron2015ocean}, the transport of oceanic pollutants such as microplastics \citep{zhang2017transport,mountford2019eulerian,van2020plastic}, and the blooming of phytoplankton \citep{durham2013turbulence,lindemann2017dynamics}. When the surface is significantly deformed or broken, strong energy exchanges take place between the turbulence in the bulk and the free surface \citep{brocchini2001dynamics,deike2022mass}. Even when the deformation of the latter is negligibly small, however, the dynamics is highly complex \citep{magnaudet2003high}. Here we focus on such a regime, considering the fundamental case in which the turbulence below the quasi-flat free surface is approximately homogeneous and isotropic. In particular, we focus on the fine-scale structure, topological properties and dynamics of the surface flow.

The study of free-surface turbulence can be traced back to \cite{uzkan1967shear} and \cite{thomas1977grid} who investigated grid turbulence adjacent to a solid wall moving at the same velocity as the mean flow. Those experimental studies showed that the surface-normal velocity fluctuations decay to vanishingly small levels over a near-wall region (later termed source layer) whose thickness is roughly one integral length scale. Following these works, \cite{hunt1978free} proposed a theoretical framework based on rapid distortion theory (RDT), describing the inviscid response of homogeneous and isotropic turbulence (HIT) to the insertion of an impermeable surface. They identified two layers: the source layer, and a viscous layer where the shear stress along the wall is brought to zero. Their predictions compared favourably with free-surface turbulence experiments in stirred tanks \citep{brumley1987near,variano2013turbulent} as well as direct numerical simulations (DNS) \citep{walker1996shear,shen1999surface,guo2010interaction,herlina2014direct} and large eddy simulations \citep{calmet2003statistical}. \cite{perot1995shear} gave a different interpretation of the interaction between the turbulence in the bulk and the non-deformable free surface, proposing that the imbalance between upwellings and downwellings (carrying fluid to and from the surface, respectively) determines the net intercomponent energy transfer. The issue was further examined by \cite{magnaudet2003high} who found that RDT is a correct leading-order approximation of the shear-free boundary layer in the limit of large Reynolds number. The latter is typically defined as \(Re_{T} = 2u'\mathcal{L}/\nu\), where \(u’\), \(\mathcal{L}\), and \(\nu\) are the root mean square (rms) velocity fluctuation, the integral scale of the turbulence in the bulk, and the kinematic viscosity, respectively. The validity of \cite{hunt1978free} theory for single-point statistics and sufficiently high \(Re_{T}\)was recently confirmed experimentally by \cite{ruth2024structure}.

The majority of the aforementioned studies focused on the evolution of the turbulence below the free surface, while less is known regarding the dynamics on the flow along the surface itself. Its topology has been explored mostly in open channel flows, both experimentally \citep{komori1989detection,kumar1998experimental,tamburrino2007free,nikora2007large}, and numerically \citep{pan1995numerical,nagaosa1999direct,lovecchio2013time,lovecchio2015upscale}. Those studies emphasized the link between the structures generated in the near-wall boundary layer and those observed along the surface. These showed similarity to two-dimensional (2D) turbulence, particularly the persistence of surface-attached vortices, as well as some evidence of an inverse energy cascade.

The flow along the surface above HIT was considered in a series of seminal papers \citep{eckhardt2001turbulence,goldburg2001turbulence,boffetta2004large,cressman2004eulerian,larkin2009power}. The authors explored features including the velocity structure functions, which were found to scale approximately as in three-dimensional (3D) turbulence, and the velocity gradients, which were highly intermittent. Moreover, they highlighted the compressible nature of the surface velocity field, leading to dense long-lived clusters of floating particles. Comparisons between computer simulations and laboratory observations were hampered by challenges associated to the free-surface boundary condition. In the simulations, the free surface was treated as a rigid lid, which \cite{shen1999surface} demonstrated could cause significant misestimation of the pressure-strain correlation even in the limit of small deformations. In the experiments, the floating particles used to image the surface flow tended to create a layer of agglomerated particles \citep{cressman2004eulerian,turney2013air}.

Simulations capturing the liquid interface above forced turbulent flows were conducted by \cite{shen1999surface} and \cite{guo2010interaction}, including regimes of low Froude number, i.e., in which the surface tension allows only small deformations. They stressed the dynamic importance of upwelling motions in connecting vortices to the free surface. There, upwellings create hairpin structures whose head dissipates rapidly in the viscous layer with the two legs connecting perpendicularly to the free surface. This suggested that upwellings lead to the increase of the number of surface-attached vortices, as later confirmed by the simulations by \cite{babiker2023vortex}. These authors found a strong correlation between the number of surface-attached vortices and the surface velocity divergence, which in turn is related to the presence of upwellings/downwellings (see \cite{guo2010interaction,ruth2024structure}). \cite{herlina2014direct,herlina2019simulation} used interface-resolving simulations to investigate the gas transfer across the surface above HIT for a range of \(Re_{T}\). They found that the increase of fine-scale structures at higher \(Re_{T}\) determines a change in the scaling of the gas transfer rate.

Another crucial aspect of the surface flow, especially relevant for the transport of contaminants, is the relative velocity and dispersion of floating particles. This was investigated by \cite{cressman2004eulerian} who found experimentally a retarded dispersion with respect to the super-diffusive regime proposed by \cite{richardson1926atmospheric}, while the latter was approximately recovered by numerical simulations. Recently, using a large-scale jet-stirred tank, we showed how the surface flow compressibility leads to anomalously large relative velocities at small separations, causing the ballistic regime of pair dispersion to extend over the inertial range of temporal separations \citep{li2024relative}. This study was the first to reach a sufficient scale separation for the emergence of the classic power-law scaling of \cite{kolmogorov1941local} theory in the surface velocity field. However, the flow was imaged by following sparse floating particles which did not allow to resolve the dissipative scales.

The above clearly indicates how, despite the importance of fine-scale flow features for a wealth of relevant processes, the detailed topology and dynamics of free-surface turbulence has not been sufficiently documented to comprehensively describe its behaviour. This is in stark contrast with 3D turbulence, for which the properties of the velocity gradient tensor and velocity differences over dissipative and inertial scales, as well as their role in the dynamics, have been explored in great depth in both the Eulerian and the Lagrangian frames \citep{sreenivasan1997phenomenology,meneveau2011lagrangian,johnson2024multiscale}. Therefore, many fundamental questions remain to be clarified: What are the spatial and temporal scales associated with the divergence, vorticity, and strain-rate of the surface flow? Does the classic scaling of velocity differences hold in free-surface turbulence? How do the upwelling and downwelling events affect the dynamics, particularly the surface vorticity and strain-rate? Addressing those and related questions is crucial, e.g., to devise effective coarse-grained representations of the surface flow, in particular considering the vast range of scales at play in nature.

Here, we conduct and analyse an experimental campaign in which the free-surface flow above homogeneous turbulence is characterized using particle tracking velocimetry (PTV). By imaging microscopic floating particles at high spatial and temporal resolution, we capture velocity gradients along dense trajectories, which allows us to gain a comprehensive view of the processes. The paper is organized as follows. In section \ref{sec_setup}, the experimental setup and methodology are introduced, and the considered flow regime is described. In section \ref{sec_result}, kinematic relations between the surface divergence, vorticity and strain-rate are derived (section \ref{sec_kinematic}); those quantities are described in terms of single-point statistics (section \ref{sec_pdf}) and structure topology (section \ref{sec_topology}). The two-point/two-time statistics are presented in terms of velocity structure functions (section \ref{sec_struct_func}), Eulerian and Lagrangian autocorrelations (section \ref{sec_scales}) and cross-correlations (section \ref{sec_cross_corr}). The clustering of floating particles is discussed in section \ref{sec_cluster}. We summarize our findings and draw conclusions in section \ref{sec_cluster}.

\section{Experimental setup and method}\label{sec_setup}

\subsection{Experimental setup}

\begin{figure}
    \centering
    \includegraphics[width=0.9\linewidth]{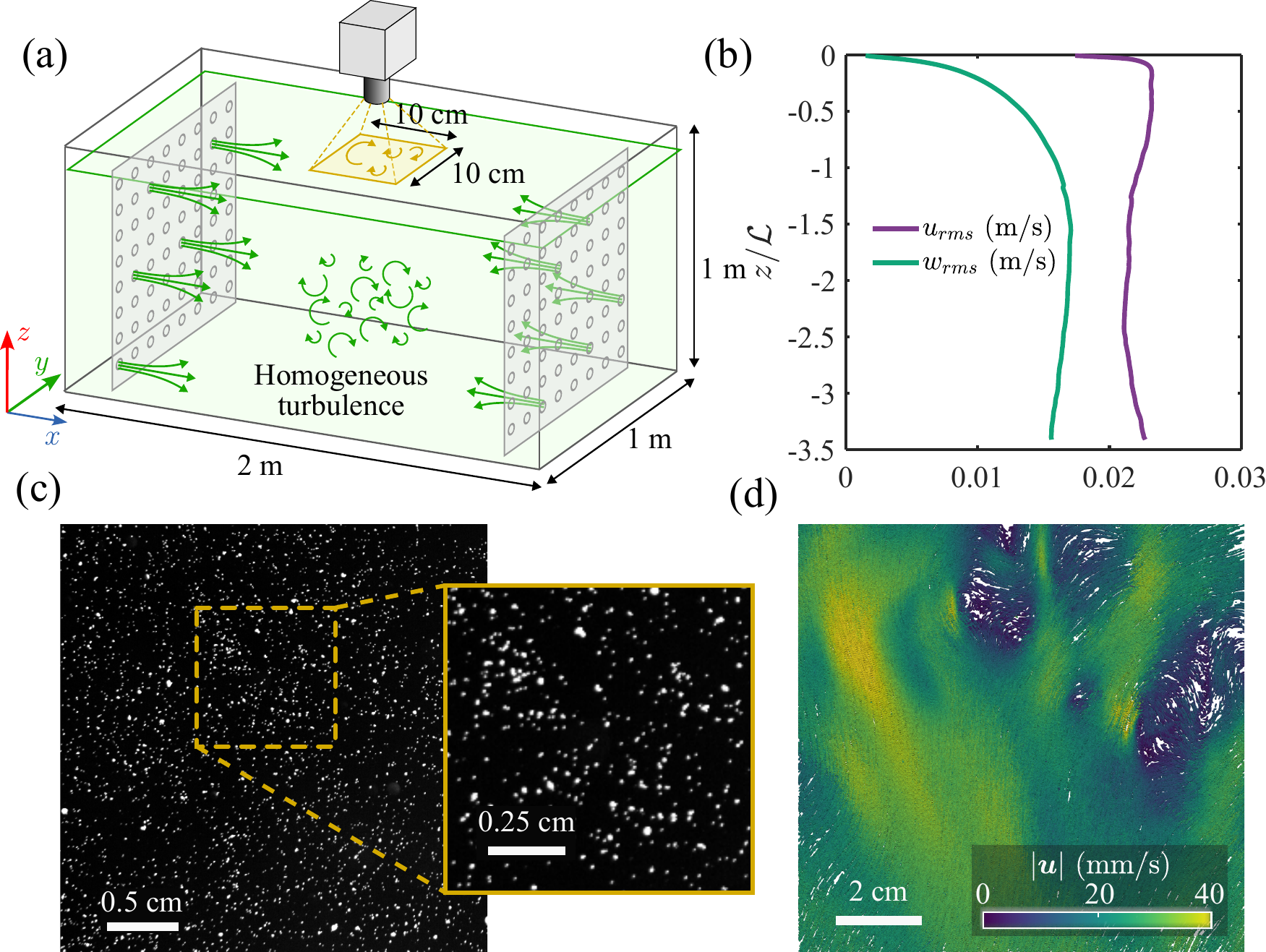}
    \caption{(a) A schematic of the turbulent water tank and camera arrangement. The yellow shaded area represents the FOV. (b) Profiles of surface-parallel and surface-normal rms fluctuation velocity (\(u_{rms}\) and \(w_{rms}\), respectively) along the vertical direction. (c) A portion of a snapshot illustrating the floating micro-particles. (d) An example of surface trajectories at \(Re_{\lambda} = 312\), color-coded by the velocity magnitude \(\left| \boldsymbol{u} \right|\).}
    \label{fig_tank}
\end{figure}

Experiments are conducted in a turbulent water tank as illustrated in Figure \ref{fig_tank}(a). The tank has dimensions of \(2 \times 1 \times 1\) m\textsuperscript{3}. In this tank, two $8\times8$ arrays of submerged pumps are placed against each other, with adjacent pumps separated by 10 cm in the horizontal and vertical directions. These pumps are controlled by programmable logic controllers and are turned on and off in a random pattern following the algorithm proposed by \cite{variano2008random}. On average, one in eight pumps is on at any given time and each jet emission lasts 3 seconds. The turbulence generated in the center of the tank is approximately homogeneous over a region of about (0.5 m)\textsuperscript{3}. The intensity of the velocity fluctuations and the dissipation rate of the turbulent kinetic energy \(\epsilon\) can be adjusted by changing the power supplied to each pump. We denote with \(x\) and \(y\) the horizontal directions parallel and perpendicular to the pump axes, respectively, and with \(z\) the vertical upward direction, the origin being at the free surface; \(u\), \(v\) and \(w\) are the respective components of the velocity vector \(\boldsymbol{u}\). Further details regarding the facility can be found in \cite{ruth2024structure} and \cite{li2024relative}.

The water level is 8 cm (which is around one integral scale) above the axis of the top row of jets. This is significantly smaller compared to most previous experiments in which the turbulence was forced at depth \citep{brumley1987near,mckenna2004role,herlina2008experiments,variano2008random,variano2013turbulent}. Therefore, as discussed in \cite{ruth2024structure}, the spatial decay of turbulence away from the forcing region is marginal and the evolution of the flow in \(z\) direction is mostly caused by the free-surface boundary condition. The latter impacts especially the surface-normal component of the velocity, as illustrated in Figure \ref{fig_tank}(b) which shows vertical profiles of surface-normal and surface-parallel rms velocity fluctuations (\(w_{rms}\) and \(u_{rms}\), respectively) obtained and described by \cite{ruth2024structure} using particles image velocimetry (PIV). During the experiment, the surface remains essentially flat, with deformation amplitude \(< 0.5\) mm as measured by planar laser-induced fluorescence \citep{ruth2024structure}. The surface is periodically skimmed to avoid accumulation of surfactants, and a surface tension of 0.07 N/s is measured using a Du Noüy ring at various points in time without seeding particles. We note that the results presented in this work are not sensitive to the exact time between the skimming and measurements, and they are robust once the free surface is recently skimmed. Still, the effect of residual surfactants is visible in the decay of the surface-parallel fluctuations approaching the surface. Similar trends were observed by \citet{variano2013turbulent} despite their efforts in cleaning the surface. Complete removal of the residual surfactant requires chemical processes; their effect, however, would not last sufficiently long time for the completion of the present measurements.

As illustrated in Figure \ref{fig_tank}(a), a downward looking CMOS camera is placed about 0.31 m above the surface to capture the surface motion within a $10\times 10$ cm\textsuperscript{2} field of view (FOV) illuminated by two LED panels. The camera has a resolution of $1664\times1600$ pixels and is operated at 400 frames per second. The fluid motion on the surface is characterized by seeding 63--75 µm floating polyethylene microspheres with a density of 0.31g/cm\textsuperscript{3}. To resolve the small-scale structures, the concentration of particles is maintained at about 120 particles/cm\textsuperscript{2}, leading to a mean inter-particle separation of around 1 mm. As the particles have a narrow size distribution and their mutual distance is much larger than their diameter, aggregation is minimized and individual particles can be clearly identified and tracked (Figure \ref{fig_tank}(c)). This is done using an in-house PTV code based on the nearest-neighbour algorithm \citep{petersen2019experimental}. Given the particle trajectories, the velocity is obtained by convolving the trajectories with the first derivative of a temporal Gaussian kernel. The width of the kernel is determined following the approach by \cite{mordant2004experimental}, and the resulting width (35 to 55 frames) is comparable to the smallest time scales of the flow. An example of trajectories in the FOV over 25 frames is shown in Figure \ref{fig_tank}(d).

\begin{figure}
    \centering
    \includegraphics[width=0.85\linewidth]{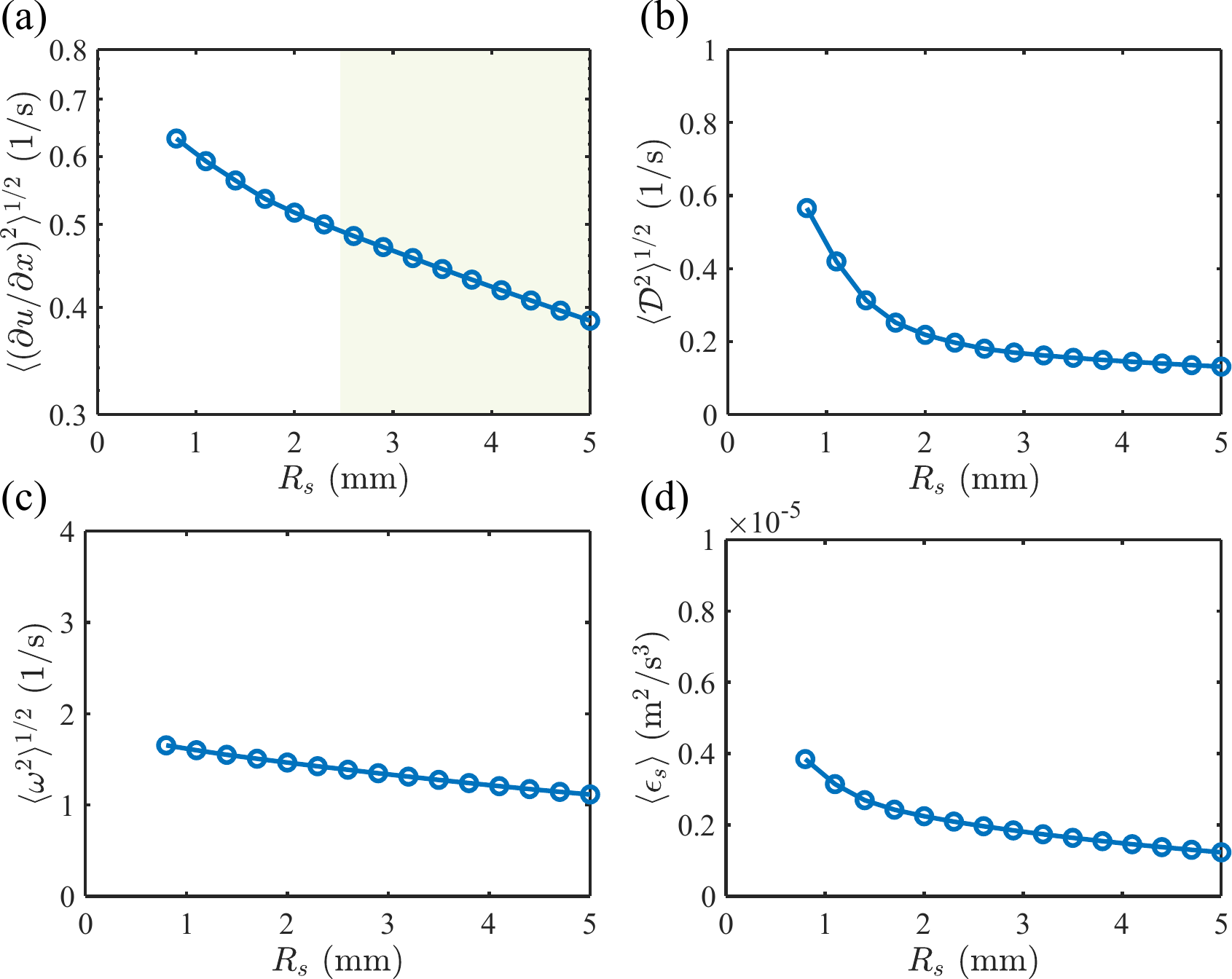}
    \caption{The evolution of different quantities as a function of the search radius \(R_{s}\): (a) the standard deviation of one component of velocity gradient tensor \(\langle \left( \partial u/\partial x \right)^{2}\rangle^{1 \slash 2}\); (b) the standard deviation of divergence \(\left\langle \mathcal{D}^{2} \right\rangle^{1 \slash 2}\); (c) the standard deviation of vorticity \(\left\langle \omega^{2} \right\rangle^{1 \slash 2}\); (d) mean dissipation rate on the surface \(\langle\epsilon_{s}\rangle\). The green shade in panel (a) marks the range of exponential decay (\(R_{s} > 2.5\) mm).}
    \label{fig_search_rad}
\end{figure}

\subsection{Velocity gradient calculation}

In order to probe the small-scale structure on the free surface, the surface velocity gradient \(\nabla_{s}\boldsymbol{u}\) is calculated, where \(\nabla_{s} = (\partial/\partial x)\boldsymbol{i} + (\partial/\partial y)\boldsymbol{j}\) is the surface gradient with \(\boldsymbol{i}\) and \(\boldsymbol{j}\) being the unit vector along \(x\) and \(y\) directions, respectively. For a given particle located at \(\boldsymbol{x}^{0}\) on the free surface, the velocity of surrounding particles located at \(\boldsymbol{x}^{p}\) within a search radius \(R_{s}\) around \(\boldsymbol{x}^{0}\) can be approximated by the leading terms in the Taylor expansion:
\begin{equation}\label{eqn_taylor}
    \boldsymbol{u}\left( \boldsymbol{x}^{p} \right) \approx \boldsymbol{u}(\boldsymbol{x}^{0}) + \nabla_{s}\boldsymbol{u}( \boldsymbol{x}^{0} )( \boldsymbol{x}^{p} - \boldsymbol{x}^{0} ),
\end{equation}

with \(p = 1,\ldots,n\). \(\nabla_{s}\boldsymbol{u}\) at \(\boldsymbol{x}^{0}\) is uniquely determined from equation \ref{eqn_taylor} when two surrounding particles are found. In the case of more than two surrounding particles, \(\nabla_{s}\boldsymbol{u}\) is calculated by minimizing the squared residuals \(\sum_{p}\left\lbrack \boldsymbol{u}\left( \boldsymbol{x}^{p} \right) - \boldsymbol{u}\left( \boldsymbol{x}^{0} \right) - \nabla_{s}\boldsymbol{u}\left( \boldsymbol{x}^{p} - \boldsymbol{x}^{0} \right) \right\rbrack^{2}\) \citep{pumir2013tetrahedron,qi2022fragmentation}. We note that large \(R_{s}\) leads to a coarse-grained velocity gradient tensor; while small \(R_{s}\) may cause larger uncertainty as only a limited number of surrounding particles can be found. \(R_{s}\) is thus selected following a similar approach to the one used to determine the width of Gaussian kernel in PTV: \(R_{s}\) is chosen as the smallest value above which the standard deviation of \(\nabla_{s}\boldsymbol{u}\) exhibits exponential decay, as shown Figure \ref{fig_search_rad}(a). Following this method, we use \(R_{s} = 2.5\) mm yielding on average 40 particles within the search radius. As shown in Figure \ref{fig_search_rad}(b--d), the key differential quantities evaluated along the surface, such as vorticity, dissipation rate and divergence, are only weakly sensitive to the exact choice of the search radius. In this work, to further minimize the uncertainty, velocity gradient calculated based on less than 5 particles are excluded from the statistics.

\subsection{Turbulence properties}

We consider four cases in which turbulence of different intensity is forced. Table \ref{tab_props} summarizes the key parameters of the turbulence in the bulk as characterized by PIV, including the Kolmogorov length scale \(\eta\) and time scale \(\tau_{\eta}\), as well as the Taylor-microscale Reynolds number $Re_\lambda$. Moreover, to illustrate to which degree the surface flow approximates HIT, we calculate from the surface PTV data the homogeneity deviation \(\text{HD} = 2\sigma_{u'}/u'\), the isotropy factor \(\text{IF} = \langle(\partial u/\partial x)/(\partial v/\partial y)\rangle\), and the mean strain-rate factor \(\text{MSRF} = \langle\left( \partial\left\langle u \right\rangle/\partial x \right)/\langle {(\partial u/\partial x - \partial\left\langle u \right\rangle/\partial x)}^{2} \rangle^{\frac{1}{2}}\rangle\). Here \(\sigma_{u'}\) is the standard deviation of the spatial field of \(u'\)  \citep{carter2016generating,esteban2019laboratory}, and angle brackets indicate ensemble averaging. The levels of HD and IF indicate a high level of spatial homogeneity and small-scale isotropy for all considered cases, while the small MSRF demonstrates that the mean velocity gradients are negligible compared to the instantaneous ones.

\begin{table}
  \begin{center}
\def~{\hphantom{0}}
  \begin{tabular}{lccccccccc}
    \(Re_{\lambda}\) & $Re_T$ & \(\epsilon\) (m\(^{2}/\)s\(^{3}\)) & \(\eta\) (mm) & \(\tau_{\eta}\) (s) & \(\mathcal{L}\) (m) & HD & IF & MSRF & \(\mathcal{C}\) \\
    207 & 2630 & \(3.82 \times 10^{- 5}\) & 0.40 & 0.16 & 0.072 & 0.26 & 1.00 & 0.056 & 0.013 \\
    248 & 3427 & \(8.31 \times 10^{- 5}\) & 0.33 & 0.11 & 0.076 & 0.24 & 1.00 & -0.047 & 0.015 \\
    283 & 4292 & \(1.44 \times 10^{- 4}\) & 0.29 & 0.08 & 0.080 & 0.21 & 0.98 & 0.064 & 0.020 \\
    312 & 5224 & \(2.21 \times 10^{- 4}\) & 0.26 & 0.07 & 0.083 & 0.18 & 1.00 & 0.012 & 0.024 \\
  \end{tabular}
  \caption{The main turbulence properties for the considered cases. The Taylor-microscale Reynolds number $Re_\lambda$, the large-scale Reynolds number $Re_T$, the dissipation rate $\epsilon$, the Kolmogorov length scale $\eta$ and time scale $\tau_\eta$, and the integral length scale $\mathcal{L}$ are evaluated in the bulk. The homogeneity deviation HD, the small-scale isotropy factor IF, the mean strain-rate factor MSRF, and the compressibility coefficient $\mathcal{C}$ are defined in the text and are evaluated on the free surface.}
  \label{tab_props}
  \end{center}
\end{table}

\section{Results}\label{sec_result}

\subsection{Kinematic relation for energy dissipation rate on the free surface}\label{sec_kinematic}

\begin{figure}
    \centering
    \includegraphics[width=0.85\linewidth]{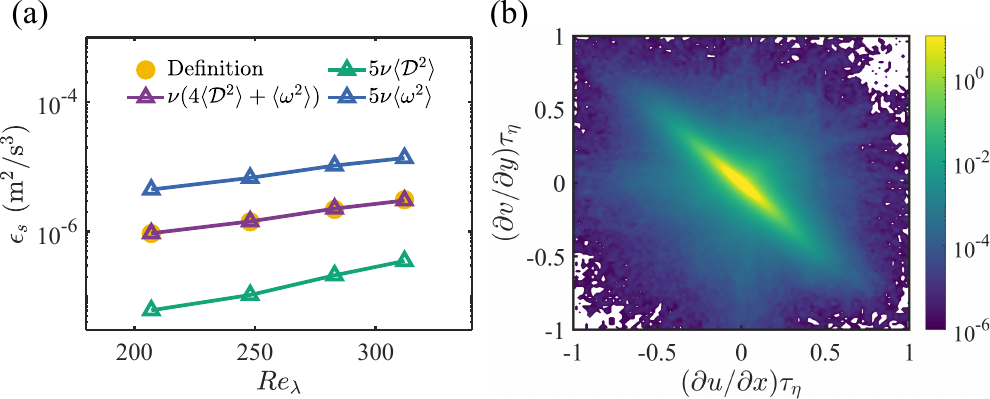}
    \caption{(a) The comparison of turbulence energy dissipation rate on the free surface \(\epsilon_{s}\) calculated based on the definition (yellow symbols), equation \ref{eqn_kinematic_eps} (purple symbols) and equation \ref{eqn_simple_eps} (blue and green symbols). (b) The joint PDF of \(\partial u/\partial x\) and \(\partial v/\partial y\) normalized by the Kolmogorov time scale \(\tau_{\eta}\) at \(Re_{\lambda} = 312\).}
    \label{fig_eps_relation}
\end{figure}

We first consider the mutual relations between vorticity, strain-rate and divergence of the surface velocity field. The surface divergence \(\mathcal{D}\) is defined as
\begin{equation}
    \mathcal{D =}\nabla_{s} \cdot \boldsymbol{u} = \partial u/\partial x + \partial v/\partial y.
\end{equation}
Considering the incompressibility of the fluid, \(\mathcal{D}\) can also be expressed by \(\mathcal{D} = - \partial w/\partial z\). Given the no-penetration boundary condition, \(w = 0\) at \(z = 0\) (which is approximately valid in the present case of weak surface deformation), positive/negative divergence represents upwelling/downwelling events. The vorticity and the strain-rate on the free surface are, respectively:
\begin{equation}
    \omega = \nabla_{s} \times \boldsymbol{u} = \partial v/\partial x - \partial u/\partial y,
\end{equation}
and
\begin{equation}
    s = \sqrt{\boldsymbol{S}_{s}\boldsymbol{S}_{s}}.
\end{equation}
where \(\boldsymbol{S}_{s} = \left\lbrack \nabla_{s}\boldsymbol{u} + \left( \nabla_{s}\boldsymbol{u} \right)^{T} \right\rbrack/2\) is the symmetric 2 by 2 strain-rate tensor associated to the 2D velocity field along the surface.

To connect the surface dynamics with the local properties of turbulence, the energy dissipation rate on the free surface is also examined, i.e., \(\epsilon_{s} = 2\nu\left\langle \boldsymbol{SS} \right\rangle.\) Note that here \(\boldsymbol{S} = \left\lbrack \nabla\boldsymbol{u} + \left( \nabla\boldsymbol{u} \right)^{T} \right\rbrack/2\) is the full 3 by 3 strain-rate tensor. As \(\epsilon_{s}\) is evaluated along the free surface, the boundary conditions allow significant simplifications. In particular, considering the free surface is quasi-flat, \(w\) is identically zero along the surface. This leads to \(\partial w/\partial x = \partial w/\partial y = 0\). Also, the zero-stress boundary condition imposes \(\partial u/\partial z = \partial v/\partial z = 0\). Therefore the (1, 2), (1, 3), (2, 1), and (3, 1) components of both \(\nabla\boldsymbol{u}\) and \(\boldsymbol{S}\) are zero. It follows that \(\epsilon_{s}\) can be written as:
\begin{equation}\label{eqn_define_eps}
    \epsilon_{s} = 2\nu\left( \langle\boldsymbol{S}_{s}\boldsymbol{S}_{s}\rangle + \langle\mathcal{D}^{2}\rangle \right) = 2\nu\left( \langle s^{2}\rangle + \langle\mathcal{D}^{2}\rangle \right).
\end{equation}

We note that in the current experiment, due to residual surfactant after skimming the surface (as mentioned above), the zero-stress boundary condition might not be strictly achieved. Therefore, $\epsilon_s$ calculated based on equation \ref{eqn_define_eps} might be weaker compared to a surface completely devoid of surfactants. Assessing this deviation, however, is difficult based on the surface PTV and is beyond the scope of this study. Equation \ref{eqn_define_eps} can be further expanded and rewritten as the summation of the quadratic terms of velocity gradient:
\begin{equation}\label{eqn_eps_expand}
    \epsilon_{s} = 2\nu\left\langle 4\left( \frac{\partial u}{\partial x} \right)^{2} + \left( \frac{\partial u}{\partial y} \right)^{2} + 3\frac{\partial u}{\partial x}\frac{\partial v}{\partial y} \right\rangle.
\end{equation}
Here, considering the properties listed in table \ref{tab_props}, we have assumed the surface turbulence to be small-scale isotropic, which implies \(\partial u/\partial x = \partial v/\partial y\) and \(\partial u/\partial y = \partial v/\partial x\), and homogeneous, which implies \(\langle(\partial u/\partial x)(\partial v/\partial y)\rangle = \langle(\partial u/\partial y)(\partial v/\partial x)\rangle\) (see equation 16 in \cite{george1991locally}). Those assumptions also allow us to write:
\begin{equation}\label{eqn_div_expand}
    \mathcal{D}^{2} = 2\left( \frac{\partial u}{\partial x} \right)^{2} + 2\frac{\partial u}{\partial x}\frac{\partial v}{\partial y},
\end{equation}
\begin{equation}\label{eqn_vor_expand}
    \omega^{2} = 2\left( \frac{\partial u}{\partial y} \right)^{2} - 2\frac{\partial u}{\partial x}\frac{\partial v}{\partial y}.
\end{equation}
By comparing equations \ref{eqn_eps_expand}, \ref{eqn_div_expand}, and \ref{eqn_vor_expand}, it is evident that \(\epsilon_{s}\) can be rewritten following:
\begin{equation}\label{eqn_kinematic_eps}
    \epsilon_{s} = \nu\left( 4\langle\mathcal{D}^{2}\rangle + \langle\omega^{2}\rangle \right).
\end{equation}

This kinematic relation, which allows expressing the dissipation rate along the surface from the strength of the divergence and vorticity on it, highlights the importance of the non-solenoidal nature and surface attached vortices to the local properties of free-surface turbulence. In the case of vanishing divergence, this relation becomes the energy dissipation rate in incompressible 2D turbulence \(\epsilon_{s} = \nu\left\langle \omega^{2} \right\rangle\). Equation \ref{eqn_kinematic_eps} agrees well with the present data for all considered cases, as shown in Figure \ref{fig_eps_relation}(a). The surface dissipation rate is found to be far smaller than the bulk value \(\epsilon\). This is consistent with previous theoretical and numerical studies \citep{teixeira2000dissipation,guo2010interaction} in which a significant decrease of dissipation at the surface was found. The surface dissipation rate will be discussed further in section \ref{sec_struct_func}.

Equation \ref{eqn_kinematic_eps} could be further simplified by assuming the compressibility ratio \(\mathcal{C} = \langle\left( \nabla_{s} \cdot \boldsymbol{u} \right)^{2}\rangle/\langle\left( \nabla_{s}\boldsymbol{u} \right)^{2}\rangle = \langle\mathcal{D}^{2}\rangle/\langle\left( \nabla_{s}\boldsymbol{u} \right)^{2}\rangle \approx 0.5\) as found in previous studies. Since this can also be expressed as \(\mathcal{C = }\langle\mathcal{D}^{2}\rangle/\lbrack 2\langle\left( \partial u/\partial x - \partial v/\partial y \right)^{2}\rangle\rbrack\) in the case of homogeneous and isotropic turbulence, the condition \(\mathcal{C} = 0.5\) is equivalent to a negligibly small correlation between \(\partial u/\partial x\) and \(\partial v/\partial y\) along the surface, i.e., \(\left| \langle(\partial u/\partial x)(\partial v/\partial y)\rangle \right| \ll \langle(\partial u/\partial x)^{2}\rangle\) \citep{cressman2004eulerian,boffetta2004large}. If this is assumed, the cross-product terms in equations \ref{eqn_eps_expand}, \ref{eqn_div_expand}, and \ref{eqn_vor_expand} are dropped and equation \ref{eqn_kinematic_eps} simplifies to:
\begin{equation}\label{eqn_simple_eps}
    \epsilon_{s} = 5\nu\langle\mathcal{D}^{2}\rangle = 5\nu\langle\omega^{2}\rangle.
\end{equation}

Figure \ref{fig_eps_relation}(a), however, indicates that the data deviates considerably from this relationship. Indeed, the observed compressibility ratio (as reported in table \ref{tab_props}) is much smaller than 0.5, which in turn is rooted in a strong correlation between \(\partial u/\partial x\) and \(\partial v/\partial y\). This is clearly illustrated in figure Figure \ref{fig_eps_relation}(b), which displays the joint probability density function (PDF) of \(\partial u/\partial x\) and \(\partial v/\partial y\) for the case \(Re_{\lambda} = 312\), demonstrating strong anti-correlation between both quantities. This strong anti-correlation and the small compressibility ratio (as well as the weak surface divergence, which will be discussed in the following sections) might be influenced by residual surfactants on the free surface. The role of the latter, even in skimmed surface, was previously explored by \citet{turney2013air}. Here and in the following, this Reynolds number will be used as exemplary case, and the behavior of the other cases is analogous.

\subsection{Divergence, vorticity and strain-rate}\label{sec_pdf}

\begin{figure}
    \centering
    \includegraphics[width=0.85\linewidth]{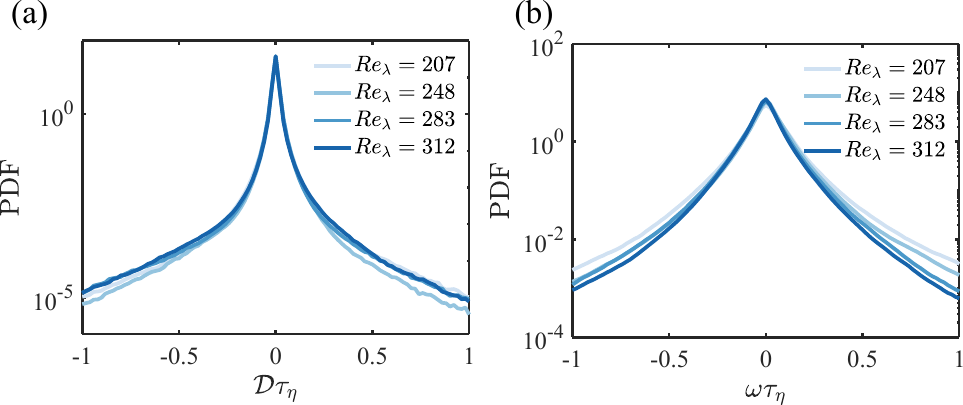}
    \caption{The PDFs of surface divergence \(\mathcal{D}\) (panel (a)) and vorticity \(\omega\) (panel (b)) at various \(Re_{\lambda}\). In both panels, darker color represents higher \(Re_{\lambda}\) and vice versa.}
    \label{fig_div_vor_pdf}
\end{figure}

Here we examine the statistical distributions of the main velocity-gradient-based quantities characterizing the surface flow: divergence, vorticity, and strain-rate. Figure \ref{fig_div_vor_pdf}(a) shows the PDF of divergence \(\mathcal{D}\) non-dimensionalized by \(\tau_{\eta}\) for the different \(Re_{\lambda}\). The symmetric distributions indicate the upwellings (associated to \(\mathcal{D >}0\)) and the downwellings (\(\mathcal{D <}0\)) occur with similar frequency and strength. The long tails signal strong intermittency, as previously observed \citep{schumacher2002clustering,cressman2004eulerian}. In addition, the approximate collapse of the PDFs for the different \(Re_{\lambda}\) suggests that the statistical behaviour of the divergence follows a dissipative scaling. This is the case also for the PDFs of \(\omega\) (Figure \ref{fig_div_vor_pdf}(b)) which however display a far greater variance, i.e., \(\langle\omega^{2}\rangle \gg \left\langle \mathcal{D}^{2} \right\rangle.\) The relatively small magnitude of \(\mathcal{D}\) is consistent with the small values of the compressibility coefficient as discussed above. We remark that all components of the velocity gradient tensor display symmetric distributions. This is in contrast with 3D turbulence, where the skewness of the longitudinal velocity differences is associated with the direct energy cascade \citep{davidson2015turbulence}.

\begin{figure}
    \centering
    \includegraphics[width=0.85\linewidth]{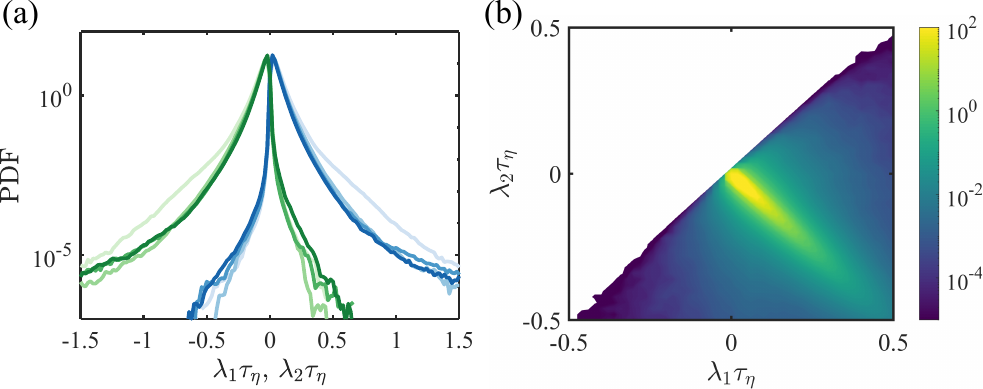}
    \caption{(a) The PDFs of the normalized eigenvalues \(\lambda_{1}\) (blue lines) and \(\lambda_{2}\) (green lines) of \(\boldsymbol{S}_{s}\) at various \(Re_{\lambda}\). Here, darker color represents higher \(Re_{\lambda}\) and vice versa. (b) The joint PDF of \(\lambda_{1}\) and \(\lambda_{2}\) for \(Re_{\lambda} = 312\).}
    \label{fig_strain_eigen_pdf}
\end{figure}

To examine the strain-rate, Figure \ref{fig_strain_eigen_pdf}(a) shows the PDFs of the eigenvalues \(\lambda_{1}\) and \(\lambda_{2}\) of \(\boldsymbol{S}_{s}\), with \(\lambda_{1} > \lambda_{2}\). As \(\boldsymbol{S}_{s}\) is a $2\times 2$ symmetric tensor, both $\lambda_1$ and $\lambda_2$ are real numbers. The distributions of both eigenvalues are clearly antisymmetric. We remind that in 3D turbulence, two out of three eigenvalues tend to be positive which indicates bi-axial stretching \citep{betchov1956inequality,davidson2015turbulence}. \cite{cardesa2013invariants} investigated the reduced strain-rate tensor and the two associated eigenvalues along 2D sections of 3D turbulence, finding predominance of compression over stretching. Along the free surface, on the other hand, compression and stretching appear equally likely and intense, similarly as the instances of positive and negative divergence (see Figure \ref{fig_div_vor_pdf}(a)). The collapse of \(\lambda_{1}\) and \(\lambda_{2}\) for different \(Re_{\lambda}\) indicates that Kolmogorov scaling again applies, as for \(\mathcal{D}\) and \(\omega\). The structure of the strain field is further clarified by the joint PDF of both eigenvalues displayed in Figure \ref{fig_strain_eigen_pdf} (b), the other Reynolds numbers showing analogous behaviour. The strong anti-correlation indicates a high likelihood of \(\lambda_{1} \approx -\lambda_{2}\), i.e., comparable strength of compression and stretching along perpendicular directions. This is consistent with Figure \ref{fig_eps_relation}(b) displaying relatively small surface divergence, which can be expressed as \(\mathcal{D}=\lambda_{1} + \lambda_{2}\).

\begin{figure}
    \centering
    \includegraphics[width=0.85\linewidth]{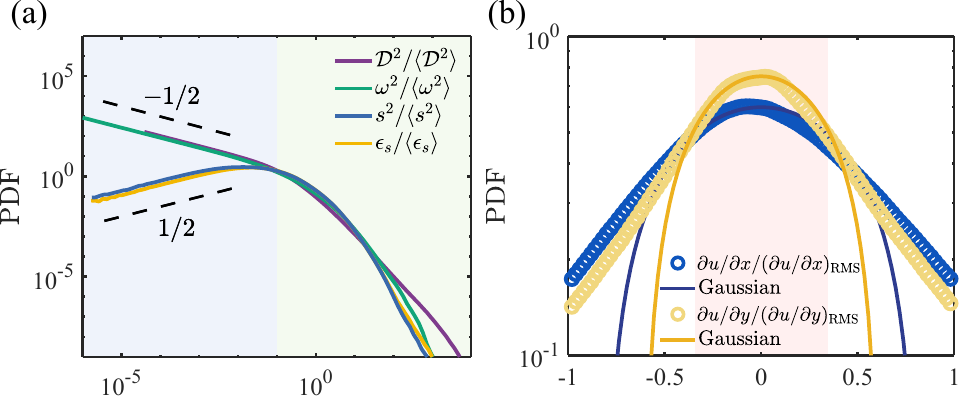}
    \caption{(a) The PDFs of normalized surface divergence square \(\mathcal{D}^{2}\), vorticity square \(\omega^{2}\), strain-rate square \(s^{2}\), and energy dissipation rate on the free surface \(\epsilon_{s}\). The dashed lines mark the scaling of power-law tails. The blue and green shaded area illustrate the region where these quantities are smaller and larger than 10\% of their mean values, respectively. (b) The PDFs of two components of velocity gradient tensor \(\partial u/\partial x\) (yellow symbols) and \(\partial u/\partial y\) (blue symbols) normalized by their rms. The solid lines show the fitted Gaussian distribution. The red shaded area from -0.3 to 0.3 marks the region where the PDFs are approximately Gaussian.}
    \label{fig_square_pdf}
\end{figure}

The magnitude of the different quantities is compared in Figure \ref{fig_square_pdf}(a), showing the PDFs of \(\mathcal{D}^{2}\), \(\omega^{2}\), \(s^{2}\) and \(\epsilon_{s}\), normalized by their mean value. In the range of low intensity events (blue shaded area), the distributions show power-law scaling with slopes of \(1/2\) for \(s^{2}\) and \(\epsilon_{s}\), and \(- 1/2\) for \(\mathcal{D}^{2}\) and \(\omega^{2}\). Power-law tails over the small-magnitude range were also observed for PDFs of squared vorticity and strain-rate in 3D turbulence by \cite{yeung2012dissipation} and \cite{carter2018small}, who explained them by the ansatz that small-velocity-gradient events behave as random variables. This is also the case here, as illustrated by Figure \ref{fig_square_pdf}(b) where PDFs of \(\partial u/\partial x\) and \(\partial u/\partial y\) are shown. It is evident that both quantities (as other components of the velocity gradient, not shown) approximately follow a Gaussian distribution when their magnitude is relatively small, e.g., less than 30\% of their rms values as indicated in Figure \ref{fig_square_pdf}(b). As the quantities in Figure \ref{fig_square_pdf}(a) are summation of squares of velocity gradient components, we expect them to follow chi-square distributions:
\begin{equation}
    P(X) \sim X^{k/2 - 1}e^{- X/2},
\end{equation}
where \(X\) is the variable representing \(\mathcal{D}^{2}\), \(\omega^{2}\), \(s^{2}\), and \(\epsilon_{s}\), \(k\) is the order of chi-square distribution specifying the number of independent squared terms being summed, and \(e\) is the natural exponent. For small \(X\), this yields a power-law tail with a slope of \(k/2 - 1\). For \(\mathcal{D}^{2}\) and \(\omega^{2}\), only one squared term is involved with \(k = 1\), and the power-law slope \(- 1/2\) is retrieved. The squared strain-rate can be written as \(s^{2} = (\partial u/\partial x)^{2} + (\partial v/\partial y)^{2} + (\partial u/\partial y + \partial v/\partial x)^{2}/2\), hence \(k = 3\) which yields the observed \(1/2\) slope. Finally, the surface dissipation can be expressed by \(\epsilon_{s} = 2\nu\left\lbrack s^{2} + (\partial w/\partial z)^{2} \right\rbrack\), where \(\partial w/\partial z\) is not an independent term considering the incompressibility condition. Therefore, \(k = 3\) is again obtained, and the scaling of the low-range tail of \(\epsilon_{s}\) follows the one of \(s^{2}\).

At the opposite end (green shaded area in Figure \ref{fig_square_pdf}(a)), we notice that the right tails of the PDFs of \(\omega^{2}\) and \(s^{2}\) follow similar patterns. This was also observed in 3D turbulence \citep{yeung2012dissipation}, suggesting that intense events of strain and vorticity are concurrent. The distribution of \(\epsilon_{s}\) essentially matches that of \(s^{2}\), consistent with equation \ref{eqn_define_eps} which results in \(\epsilon_{s} \approx 2\nu s^{2}\) for \(\mathcal{C} \ll 1\). Though the divergence is in general relatively small, its intermittency is even higher than the other analysed quantities. The overall strong intermittency of the velocity gradient as well as its associated quantities on the free surface was recently found to be associated with the nonlinear self-amplification of the velocity gradient \citep{qi2024restricted}, which also accounts for the strong intermittency in 3D turbulent flows \citep{meneveau2011lagrangian,johnson2024multiscale}.

\subsection{Topology of small-scale structures}\label{sec_topology}

\begin{figure}
    \centering
    \includegraphics[width=0.95\linewidth]{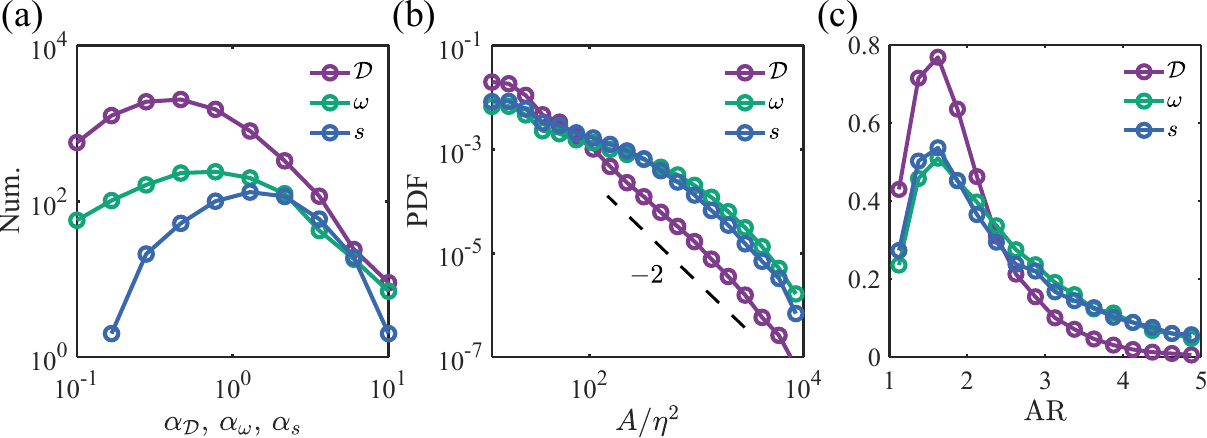}
    \caption{(a) The number of high-intensity objects found in the FOV as a function of thresholds. (b) PDFs of the normalized area of high-intensity objects. The dashed line marks the power-law scaling of \(- 2\). (c) PDFs of the aspect ratio of high-intensity events. In all the panels, the purple, green and blue symbols represent high-divergence, high-vorticity and high-strain objects, respectively. Only the data for \(Re_{\lambda} = 312\) is included.}
    \label{fig_topology}
\end{figure}

We then examine the topology of small-scale structures; in particular, three sets of discrete structures used to characterize the spatial organization of events of high surface-divergence, vorticity, and strain-rate. Those structures are defined as contiguous regions satisfying the conditions \(\left| \mathcal{D} \right| > \alpha_{\mathcal{D}}{\langle\mathcal{D}^{2}\rangle}^{1/2}\), \(|\omega| > \alpha_{\omega}{\langle\omega^{2}\rangle}^{1/2}\), and \(s > \alpha_{s}\langle s\rangle\), where \(\alpha_{\mathcal{D}}\), \(\alpha_{\omega}\), and \(\alpha_{s}\) are positive constants. To determine appropriate thresholds, we analyse the percolation behaviour of the intense structures as first proposed by \cite{moisy2004geometry}. For high threshold values, only a few small objects can be detected. As the threshold decreases, the objects grow in size and number and eventually start merging. The optimal threshold is obtained by identifying the intermediate value for which the objects are most numerous. This procedure was used extensively to identify structures in various configurations including channel flows \citep{lozano2012three}, free shear flows \citep{dong2017coherent}, and homogeneous turbulence \citep{carter2018small}.

Here, the velocity gradient measured at the position of each particle is first interpolated on a Cartesian grid with size equal to half the mean inter-particle distance. Figure \ref{fig_topology}(a) then shows how the number of detected objects varies as a function of threshold level, yielding to the choice \(\alpha_{\mathcal{D}} = 0.4\), \(\alpha_{\omega} = 0.6\), and \(\alpha_{s} = 1.5\). It is noted that the following results are not sensitive to the exact values of thresholds. Moreover, objects that touch the FOV boundary are discarded. Although this may lead to underestimating the number of large structures, it will be shown that the vast majority of the identified objects are much smaller than the FOV.

Figure \ref{fig_topology}(b) shows the PDFs of the area of high-divergence, high-vorticity, and high-strain-rate structures, normalized by the Kolmogorov scale. The size of the structures is widely distributed over four decades. The high-vorticity and high-strain objects follow a similar trend, confirming the correlation between events of intense \(\omega\) and \(s\). These structures are on average larger compared to the regions of high divergence. Over some size range, the distributions appear compatible with a power-law decay, which may suggest a link with the scale-invariant properties of turbulence \citep{sreenivasan1991fractals,moisy2004geometry,carter2018small}. The limited range of scales over which this is evident, however, does not allow any conclusive statement in this sense.

In order to characterize the geometry of these structure, we also consider their aspect ratio \(\text{AR} = R_{1}/R_{2}\), where \(R_{1}\) and \(R_{2}\) are the major and minor axis of an ellipse that has the same second central moments as the structure. To ensure the accurate AR calculation, objects with area smaller than 5 grid cells (corresponding to around \(5\eta^{2}\)) are not considered in these statistics. It is found the results do not display discernible dependence on the cutoff value between 3 and 9 grid cells. Figure \ref{fig_topology}(c) shows the PDFs of AR for the three types of structures. Again, the curves for the high-vorticity and high-strain objects largely overlap. Those structures have generally larger AR, indicating that high-vorticity and high-strain-rate structures are more elongated compared to those of high-divergence.

\begin{figure}
    \centering
    \includegraphics[width=0.95\linewidth]{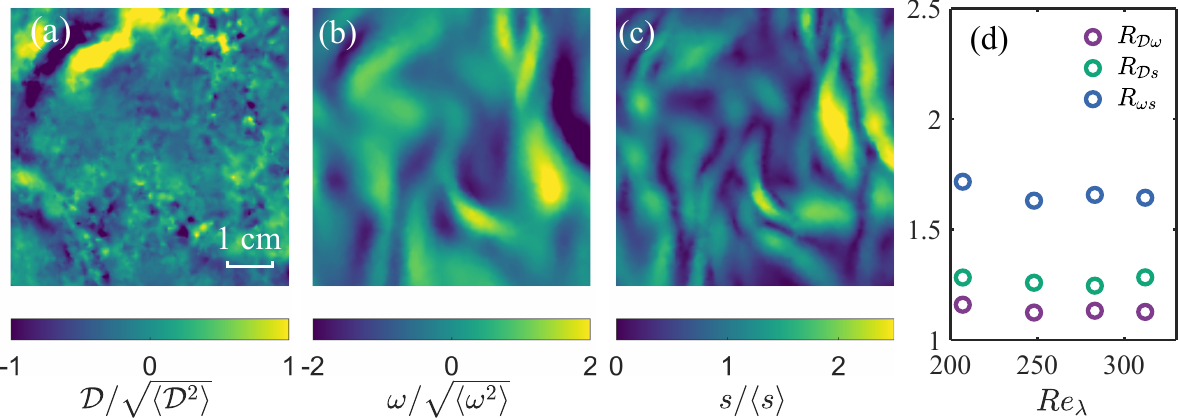}
    \caption{(a--c) Snapshots of surface divergence field (a), vorticity field (b) and strain field(c) on the free surface at \(Re_{\lambda} = 312\). (d) Overlap coefficients as a function of \(Re_{\lambda}\). The purple, green and blue symbols represent divergence-vorticity, divergence-strain and vorticity-strain overlap, respectively.}
    \label{fig_vel_grad_field}
\end{figure}

These properties are confirmed by the visual observations of instantaneous fields, samples of which are reported in Figure \ref{fig_vel_grad_field}(a--c): the high-divergence events are relatively small-scale and spotty whereas the high-vorticity and high-strain regions are larger and more elongated. This divergence snapshot is consistent with the numerical simulation at a lower Reynolds number ($Re_T\approx1800$) by \citet{herlina2019simulation}, in which the surface divergence also appears to have a smaller length scale compared with the integral scale in the bulk. Moreover, the vorticity and strain-rate fields follow similar patterns, with high-vorticity magnitude (both positive and negative) events also overlapping with high-strain regions. This concurrence of intense vorticity and strain is also found in 3D turbulence \citep{yeung2015extreme}. To quantify the topological connection between such objects, we define the overlapping coefficients between \(\mathcal{D}\), \(\omega\), and \(s\), based on a procedure similar to the one used by \cite{berk2023dynamics}. For example, the overlap between high-vorticity and high-divergence structures is characterized by:
\begin{equation}
    R_{\mathcal{D}\omega} = \frac{A_{FOV}\langle A_{\mathcal{D}\omega}\rangle}{\langle A_{\mathcal{D}}A_{\omega}\rangle},
\end{equation}
where \(A_{FOV}\) is the total area of FOV. \(A_{\mathcal{D}}\) and \(A_{\omega}\) are the area of high-divergence and high-vorticity regions, respectively. \(A_{\mathcal{D}\omega}\) denotes the overlapping area between these regions. If both type of structures are spatially uncorrelated, \(R_{\mathcal{D}\omega} = 1\). Similarly, the divergence/strain-rate and vorticity/strain-rate overlapping coefficients are defined as:
\begin{equation}
    R_{\mathcal{D}s} = \frac{A_{FOV}\langle A_{\mathcal{D}s}\rangle}{\langle A_{\mathcal{D}}A_{s}\rangle},
\end{equation}
\begin{equation}
    R_{\omega s} = \frac{A_{FOV}\langle A_{\omega s}\rangle}{\langle A_{\omega}A_{s}\rangle},
\end{equation}
where \(A_{s}\)represents the area of high-strain structures, and \(A_{\mathcal{D}s}\) and \(A_{\omega s}\) are overlapping area defined similarly. Figure \ref{fig_vel_grad_field}(d) shows \(R_{\mathcal{D}\omega}\), \(R_{\mathcal{D}s}\), and \(R_{\omega s}\) for the different \(Re_{\lambda}\). In all cases, high-vorticity and high-strain structures show significant correlation, consistent with the previous observations. \(R_{\mathcal{D}\omega}\) and \(R_{\mathcal{D}s}\) are much weaker than \(R_{\omega s}\), but the high-divergence events are more likely to be concurrent with high strain-rate than high vorticity.

\subsection{Second-order velocity structure functions}\label{sec_struct_func}

\begin{figure}
    \centering
    \includegraphics[width=0.95\linewidth]{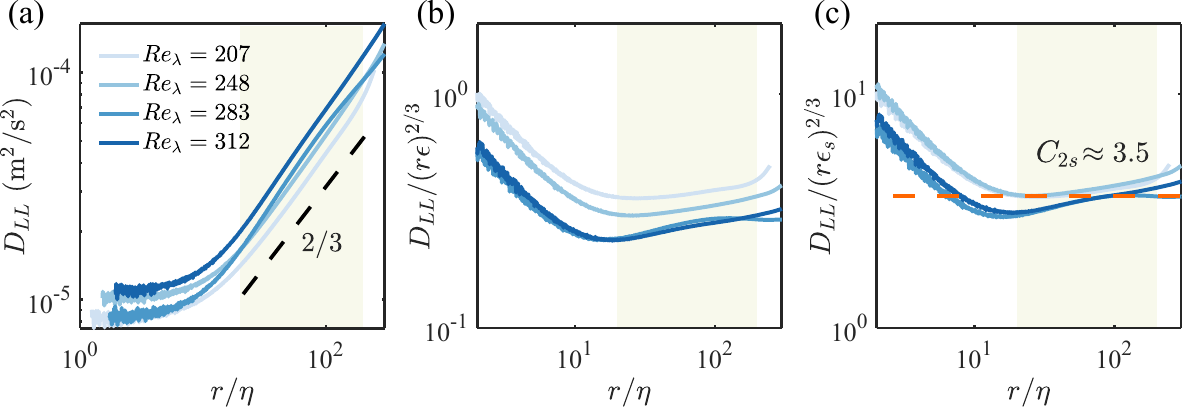}
    \caption{(a)The longitudinal second-order structure function \(D_{LL}\) as a function of the separation distance. The dashed line denotes \(2/3\) power-law scaling. (b) \(D_{LL}\) normalized by \((r\epsilon)^{2/3}\). (c) \(D_{LL}\) normalized by \(\left( r\epsilon_{s} \right)^{2/3}\). The orange dashed line marks \(C_{2s} \approx 3.5\). In all the panels, the darker color represents high \(Re_{\lambda}\) and vise versa. The green shaded area marks the inertial range.}
    \label{fig_struct_func}
\end{figure}

After exploring the small-scale properties of the surface flow field, we consider how the turbulent kinetic energy is distributed across the spatial scales as described by the second-order longitudinal structure function. This is defined as
\begin{equation}
    D_{LL} = \left\langle \left(u_{r}\left( \boldsymbol{x} + r{\hat{\boldsymbol{e}}}_{r} \right) - u_{r}\left( \boldsymbol{x} \right)\right)^{2} \right\rangle,
\end{equation}
where \(r\) is the separation distance, \({\hat{\boldsymbol{e}}}_{r}\) represents the unit vector along the separation, and \(u_{r}\) is the velocity component in the same direction. Figure \ref{fig_struct_func}(a) shows \(D_{LL}\)as a function of \(r/\eta\) for the different \(Re_{\lambda}\). In all cases, a clear scaling \(D_{LL} \sim r^{2/3}\) is visible for \(r/\eta > 20\), consistent with the classic theory of \cite{kolmogorov1941local} in the inertial sub-range. This was also observed in the laboratory experiments by \cite{goldburg2001turbulence} and \cite{cressman2004eulerian}, and in outdoor water streams by \cite{chickadel2011infrared} (who reported the equivalent scaling of the energy spectrum with \(k^{- 5/3}\), \(k\) being the wave number) and by \cite{sanness2023effect}.

At smaller separations, the structure functions do not transition to the scaling \(D_{LL}\sim r^{2}\) expected for smooth flows in the dissipation range, and instead their slope become much shallower than in the inertial sub-range. This behaviour was recently reported by \cite{li2024relative} using much sparser particle concentrations. Here the finer spatial resolution allows us not only to confirm this finding, but to reveal that the anomalously large relative velocities persist down to millimetric separations. As discussed in \cite{li2024relative}, this behaviour is similar to the formation of caustics in the velocity fields described by inertial particles in turbulence: such fields are also compressible, with intermittently large relative velocities and thus anomalous scaling exponents of the structure functions at small scales \citep{bec2010intermittency,bewley2013observation,berk2021dynamics,bec2024statistical}. We note that, as the floating particles follow the fluid motion, their relative velocity must ultimately recover the scaling \(u_{r} \sim r\) (and thus \(D_{LL} \sim r^{2}\)) in the limit of vanishing separations. This, however, may happen at scales only slightly larger than the particle diameter, not accessible even in the present high-resolution imaging system.

Besides the scaling with the separation \(r\), a crucial prediction of Kolmogorov's theory is the dependence of the structure function with the dissipation in the inertial sub-range. In 3D turbulence, the theory predicts \(D_{LL} = C_{2}(\epsilon r)^{2/3}\), where \(C_{2} \approx 2.1\) is the Kolmogorov constant \citep{sreenivasan1995universality}. The compensated plots \(D_{LL}/(\epsilon r)^{2/3}\) in Figure \ref{fig_struct_func}(b) show that this relation does not hold in free-surface turbulence: the lines are not close to the value of 2.1 in the inertial range and, more importantly, they do not collapse. In fact, the dynamics on the free surface is expected to be determined by the energy dissipation rate on the free surface \(\epsilon_{s}\) rather than by the one in the bulk water \(\epsilon\). Indeed, the compensated plots \(D_{LL}/\left( \epsilon_{s}r \right)^{2/3}\) in Figure \ref{fig_struct_func}(c) show a much better collapse which indicates that the classic scaling still holds for free-surface turbulence when the dissipation rate on the surface is considered, i.e., \(D_{LL} = C_{2s}\left( \epsilon_{s}r \right)^{2/3}\). Here, a distinct factor \(C_{2s} \approx 3.5\) is obtained.

\subsection{Scales of surface divergence, vorticity, and strain-rate}\label{sec_scales}

\begin{figure}
    \centering
    \includegraphics[width=0.95\linewidth]{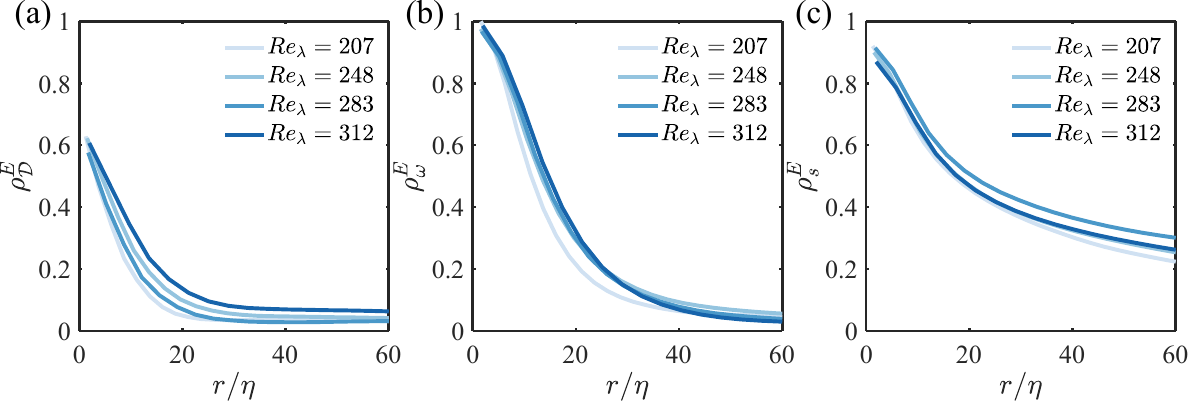}
    \caption{The Eulerian autocorrelation functions of divergence (a), vorticity (b) and strain (c) at different \(Re_{\lambda}\). In all the panel, the darker color represents higher \(Re_{\lambda}\).}
    \label{fig_euler_autocorr}
\end{figure}

As seen in section \ref{sec_topology}, the surface divergence, vorticity, and strain-rate exhibit different length scales. This aspect is further investigated by examining the spatial autocorrelation functions of each quantity:
\begin{equation}
    \rho_{\mathcal{D}}^{E} = \frac{\langle \mathcal{D}'\left( \boldsymbol{x} \right)\mathcal{D}'\left( \boldsymbol{x} + r\hat{\boldsymbol{e}}_{r} \right)\rangle}{\langle \mathcal{D}'^{2}\rangle},
\end{equation}
\begin{equation}
    \rho_{\omega}^{E} = \frac{\langle\omega'\left( \boldsymbol{x} \right)\omega'\left( \boldsymbol{x} + r{\hat{\boldsymbol{e}}}_{r} \right)\rangle}{\langle\omega'^{2}\rangle},
\end{equation}
\begin{equation}
    \rho_{s}^{E} = \frac{\langle s'\left( \boldsymbol{x} \right)s'\left( \boldsymbol{x} + r{\hat{\boldsymbol{e}}}_{r} \right)\rangle}{\langle s'^{2}\rangle},
\end{equation}
where the superscript \(E\) stands for Eulerian and the prime denotes fluctuations around the ensemble average. Figure \ref{fig_euler_autocorr} shows the autocorrelation for the various \(Re_{\lambda}\). The divergence field (Figure \ref{fig_euler_autocorr}(a)) exhibits a characteristic length scale around 5--10$\eta$, while the vorticity and strain-rate fields are characterized by somewhat larger correlation scales \(\sim 20\eta\) and \(\sim 30\eta\), respectively. This observation is consistent with the result in Figure \ref{fig_topology}(b) and Figure \ref{fig_vel_grad_field}. Beyond some experimental scatter, the autocorrelation functions show no discernible dependence on \(Re_{\lambda}\), with Kolmogorov scaling providing a fair collapse of the curves.

\begin{figure}
    \centering
    \includegraphics[width=0.95\linewidth]{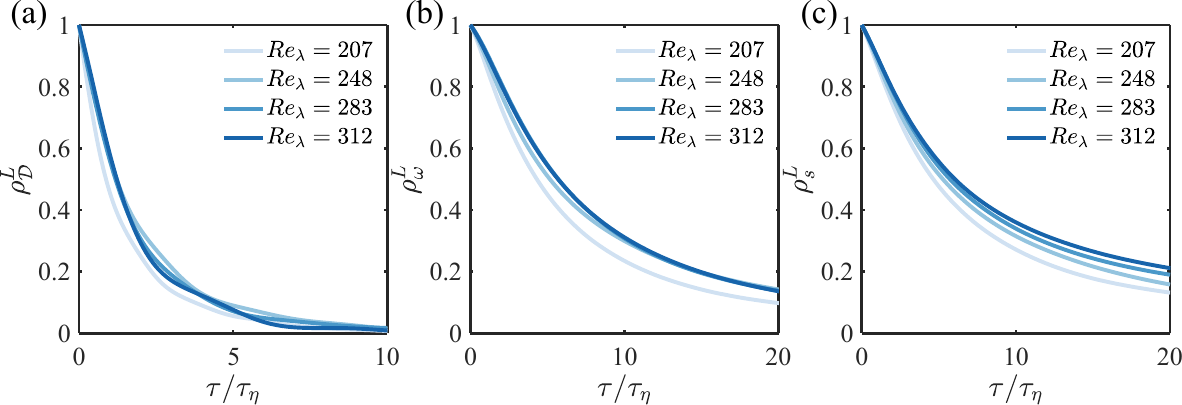}
    \caption{The Lagrangian autocorrelation functions of divergence (a), vorticity (b) and strain (c) at different \(Re_\lambda\). In all the panel, the darker color represents higher \(Re_{\lambda}\).}
    \label{fig_lag_autocorr}
\end{figure}

As the velocity gradient is obtained along floating particle trajectories, the temporal scales can also be investigated by calculating the temporal autocorrelation functions of divergence, vorticity and strain-rate, respectively:
\begin{equation}
    \rho_{\mathcal{D}}^{L} = \frac{\mathcal{\langle D'}(t)\mathcal{D'}(t + \tau)\rangle}{\mathcal{\langle D}'^{2}\rangle},
\end{equation}
\begin{equation}
    \rho_{\omega}^{L} = \frac{\langle\omega'(t)\omega'(t + \tau)\rangle}{\langle\omega'^{2}\rangle},
\end{equation}
\begin{equation}
    \rho_{s}^{L} = \frac{\langle s'(t)s'(t + \tau)\rangle}{\langle s'^{2}\rangle}.
\end{equation}
Here the superscript \(L\) stands for Lagrangian, \(t\) represents the generic temporal abscissa, and \(\tau\) is the time delay. Figure \ref{fig_lag_autocorr} shows the temporal autocorrelation functions for the different \(Re_{\lambda}\), with the Kolmogorov scaling that again provides a reasonable collapse of the curves. The surface divergence shows a time scale comparable to \(\tau_{\eta}\), confirming it is driven by small-scale processes. On the other hand, the time scales of both vorticity and strain-rate are significantly larger and comparable to the integral time scale, suggesting that surface-attached vortices and surface-parallel stretching and compression are long-lived compared to the lifetime of sources and sinks. These time scales, in fact, are also larger than their characteristic time scales in 3D turbulence, which are expected to scale with \(\tau_{\eta}\).

\begin{figure}
    \centering
    \includegraphics[width=0.7\linewidth]{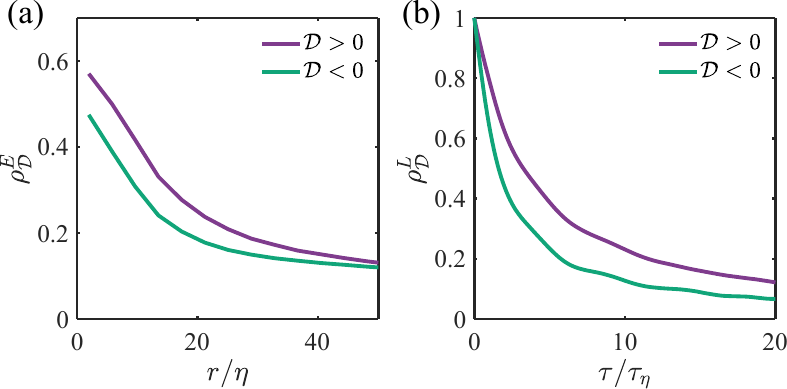}
    \caption{(a) The Eulerian autocorrelation functions for positive divergence (purple line) and negative divergence (green line). (b) The Lagrangian autocorrelation functions for positive divergence (purple line) and negative divergence (green line).}
    \label{fig_div_autocorr}
\end{figure}

Given the importance of the imbalance between upwellings and downwellings \citep{perot1995shear,guo2010interaction,ruth2024structure}, it is useful to distinguish between the scales associated to both type of events. Figure \ref{fig_div_autocorr}(a) and (b) plot the Eulerian and Lagrangian autocorrelation functions \(\rho_{\mathcal{D}}^{E}\) and \(\rho_{\mathcal{D}}^{L}\), respectively, conditioning on \(\mathcal{D} > 0\) and \(\mathcal{D} < 0\). It is clear that upwellings are associated to larger spatial and temporal scales along the surface flow compared to downwellings. This is consistent with the observation by by \cite{ruth2024structure} in the same facility and by \cite{guo2010interaction} based on numerical simulations.

\subsection{Cross-correlation among divergence, vorticity and strain-rate}\label{sec_cross_corr}

\begin{figure}
    \centering
    \includegraphics[width=0.85\linewidth]{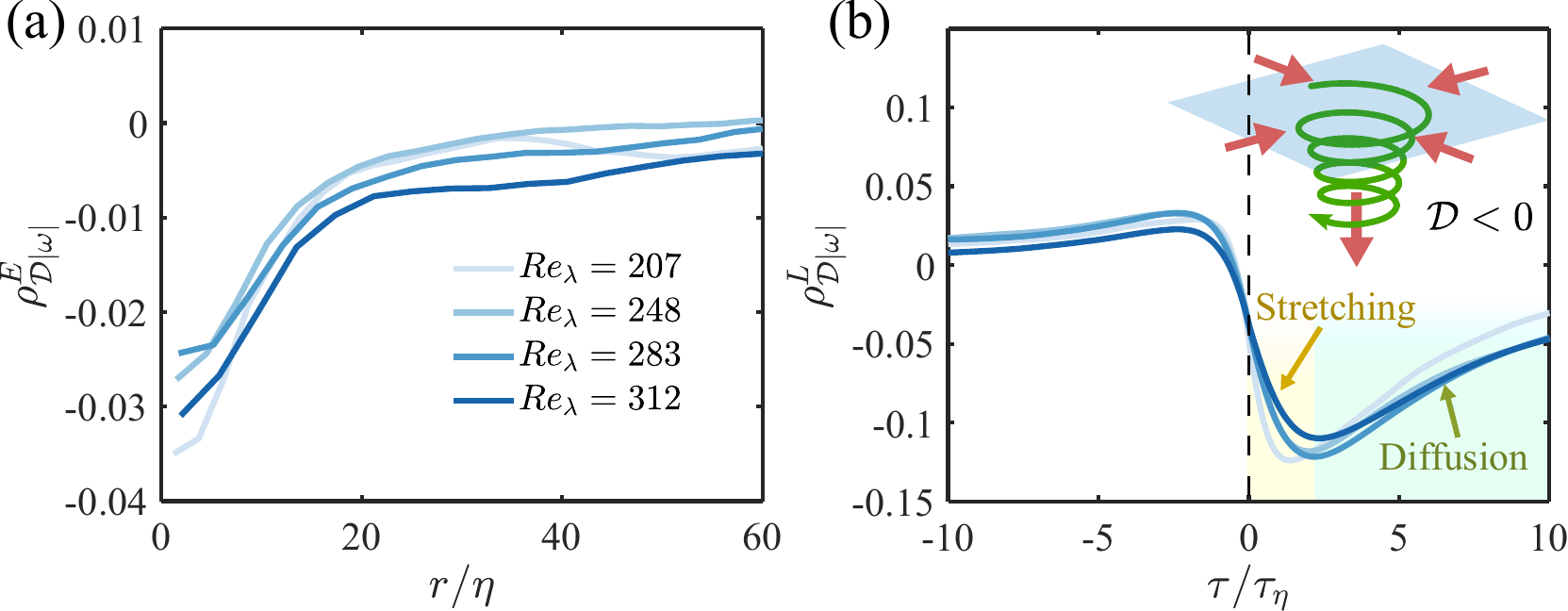}
    \caption{(a) The Eulerian cross-correlation functions between divergence and vorticity magnitude at different \(Re_{\lambda}\). (b) The Lagrangian cross-correlation function between divergence and vorticity magnitude at different \(Re_{\lambda}\). This panel shares the same legend with panel (a). The yellow shaded area marks the vortex-stretching process, and the green shaded area marks the diffusion process of surface-attached vortices. The schematic illustrates the vortex-stretching process. The green spiral marks the surface-attached vortex, the red arrows represent the negative divergence, and the blue plane shows the free surface.}
    \label{fig_div_vor_cross}
\end{figure}

Having characterized the spatial and temporal scales of divergence, vorticity and strain-rate of the surface flow, we investigate its dynamics by considering the mutual correlation between those quantities, which will prove insightful towards a mechanistic understanding of the processes. We first consider the Eulerian cross-correlation between divergence and vorticity \(\rho_{\mathcal{D}|\omega|}^{E}\)
\begin{equation}
    \rho_{\mathcal{D}|\omega|}^{E} = \frac{\langle\mathcal{D}'\left( \boldsymbol{x} \right)|\omega|'\left( \boldsymbol{x} + r{\hat{\boldsymbol{e}}}_{r} \right)\rangle}{ \mathcal{\langle D}'^{2}\rangle^{1/2}\langle|\omega|'^{2}\rangle^{1/2} },
\end{equation}
where the absolute value of the vorticity is used as the rotational direction of surface-attached vortices is immaterial. Figure \ref{fig_div_vor_cross}(a) shows the cross-correlation for various \(Re_{\lambda}\), with Kolmogorov scaling providing again a fair collapse of the different curves. For all cases, \(\rho_{\mathcal{D}|\omega|}^{E} \approx - 0.03\) at \(r = 0\) is observed. This suggests that, although the correlation between these two quantities is weak, strong vorticity is more likely to be associated with negative surface divergence; i.e., surface-attached vortices are stronger when downwelling events occur. Moreover, Figure \ref{fig_div_vor_cross}(a) indicates a characteristic correlation scale \(\approx 10\eta\). This is close to the length scale of the divergence as the latter decorrelates from itself faster than the vorticity (see Figure \ref{fig_euler_autocorr}).

To further probe the interaction between divergence and vorticity, we also examine the Lagrangian cross-correlation, which is defined as
\begin{equation}
    \rho_{\mathcal{D}|\omega|}^{L} = \frac{\mathcal{\langle D'}(t)|\omega|'(t + \tau)\rangle}{\mathcal{\langle D}'^{2}\rangle^{1/2}\langle|\omega|'^{2}\rangle^{1/2}},
\end{equation}
As shown in Figure \ref{fig_div_vor_cross}(b), this cross-correlation remains around zero before \(\tau < 0\), suggesting that strong vorticity events do not affect the divergence at later times. However, for \(\tau \geq 0\), \(\rho_{\mathcal{D}|\omega|}^{L}\) becomes negative and dips during a few Kolmogorov time units. This anti-correlation between \(\mathcal{D}\) and \(\omega\), corroborating the observation in Figure \ref{fig_div_vor_cross}(a), indicates that sinks in the surface flow are statistically associated to the enhancement of surface-attached vortices at a later time. This can be explained by considering the physical picture illustrated in the inset of Figure \ref{fig_div_vor_cross}(b): when a sink (marked by a red arrow) is formed, the correspondent downwelling flow stretches downwards surface-attached vortex filaments (marked as a green spiral). As a result, the vorticity magnitude grows and \(\rho_{\mathcal{D}|\omega|}^{L}\) decreases significantly (yellow shaded area in the plot). The typical duration of the vortex stretching, around \(2\tau_{\eta}\), is consistent with the time scale of the events of negative divergence highlighted in Figure \ref{fig_div_autocorr}(b). When the surface sink has dissipated, the vorticity diffuses and \(\rho_{\mathcal{D}|\omega|}^{L}\) approaches zero (blue shaded area). We note that this picture should hold for different \(Re_{\lambda}\) as indicated by the good collapse of the lines in Figure \ref{fig_div_vor_cross}(b).

\begin{figure}
    \centering
    \includegraphics[width=0.85\linewidth]{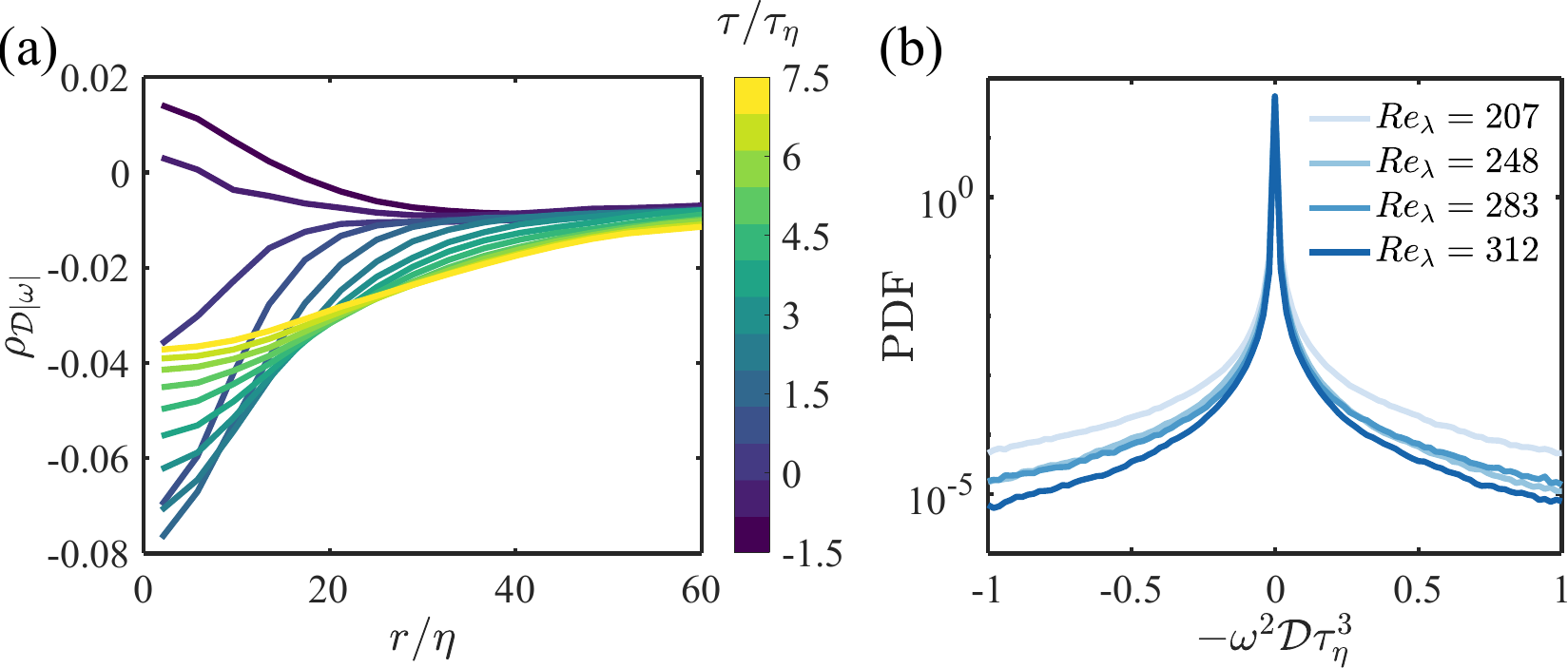}
    \caption{(a) The spatial-temporal cross-correlation between the divergence and vorticity magnitude as a function of separation distance. The colors represent the time delay \(\tau\). (b) The PDFs of the vortex-stretching term. The darker color represent the higher \(Re_{\lambda}\).}
    \label{fig_div_vor_cross_spatial_temp}
\end{figure}

This picture of the connection between vortex-stretching and surface divergence is corroborated by Figure \ref{fig_div_vor_cross_spatial_temp}(a), plotting the spatio-temporal cross-correlation between the divergence and vorticity,
\begin{equation}
    \rho_{\mathcal{D}|\omega|} = \frac{\mathcal{\langle D'}\left( \boldsymbol{x},t \right)|\omega|'\left( \boldsymbol{x} + r{\hat{\boldsymbol{e}}}_{r},t + \tau \right)\rangle}{\mathcal{\langle D}'^{2}\rangle^{1/2}\langle|\omega|'^{2}\rangle^{1/2}}.
\end{equation}
The cross-correlation is again weak for \(\tau < 0\); whereas during the initial stage of the downwelling (\(0 < \tau < 2\tau_{\eta}\)), the vorticity intensifies and the cross-correlation dips into the negative range. The anti-correlation extends spatially to \(r \approx 10\eta\), i.e., the characteristic length scale of the divergence; while after \(\tau > 2\tau_{\eta}\), when the sink dissipates, the diffusion of the vorticity leads to an increase of the cross-correlation length scale.

The connection between the surface divergence and surface-attached vortices has been explored by several authors. \cite{banerjee1994upwellings} proposed that a correlation must exist between the surface divergence and the number of surface-attached vortices, which was recently demonstrated by \cite{babiker2023vortex}. \cite{shen1999surface} used numerical simulations to provide a mechanistic interpretation of such correlation, showing how upwellings bring hairpin vortices close to the surface, where their surface-parallel section is dissipated leaving vortex filaments that connect to the surface. The present analysis has emphasized the relation between surface vortices and downwellings, rather than upwellings. The former are naturally associated to vortex stretching, which in turn is classically attributed a key role in the energy cascade. However, the picture appears to be the completely different in free-surface turbulence, as we now show.

The evolution of vorticity under the action of the strain field is often characterized by examining the vortex stretching term \(\boldsymbol{\omega} \cdot \boldsymbol{S} \cdot \boldsymbol{\omega}\) in the enstrophy transport equation, where \(\boldsymbol{\omega}\) is the vorticity vector. On the free surface, this term reduces to \(\omega^{2}(\partial w/\partial z) = - \omega^{2}\mathcal{D}\). Figure \ref{fig_div_vor_cross_spatial_temp}(b) shows the PDF of this terms normalized by Kolmogorov scales for all \(Re_{\lambda}\). The distribution is symmetric and strongly intermittent. The symmetry is in stark contrast with the behaviour of 3D turbulence, in which vortex stretching is predominant \citep{mullin2006dual,buxton2010amplification,bechlars2017variation}. The fact that the free surface acts to suppress the vortex stretching is also observed in \cite{qi2024restricted} by examining the asymmetry of the joint PDF of velocity gradient invariants. This result may provide clues regarding the energy cascade along the free surface, which has been found to show inverse energy transfer from small to large scales \citep{pan1995numerical,lovecchio2015upscale}.

\begin{figure}
    \centering
    \includegraphics[width=0.85\linewidth]{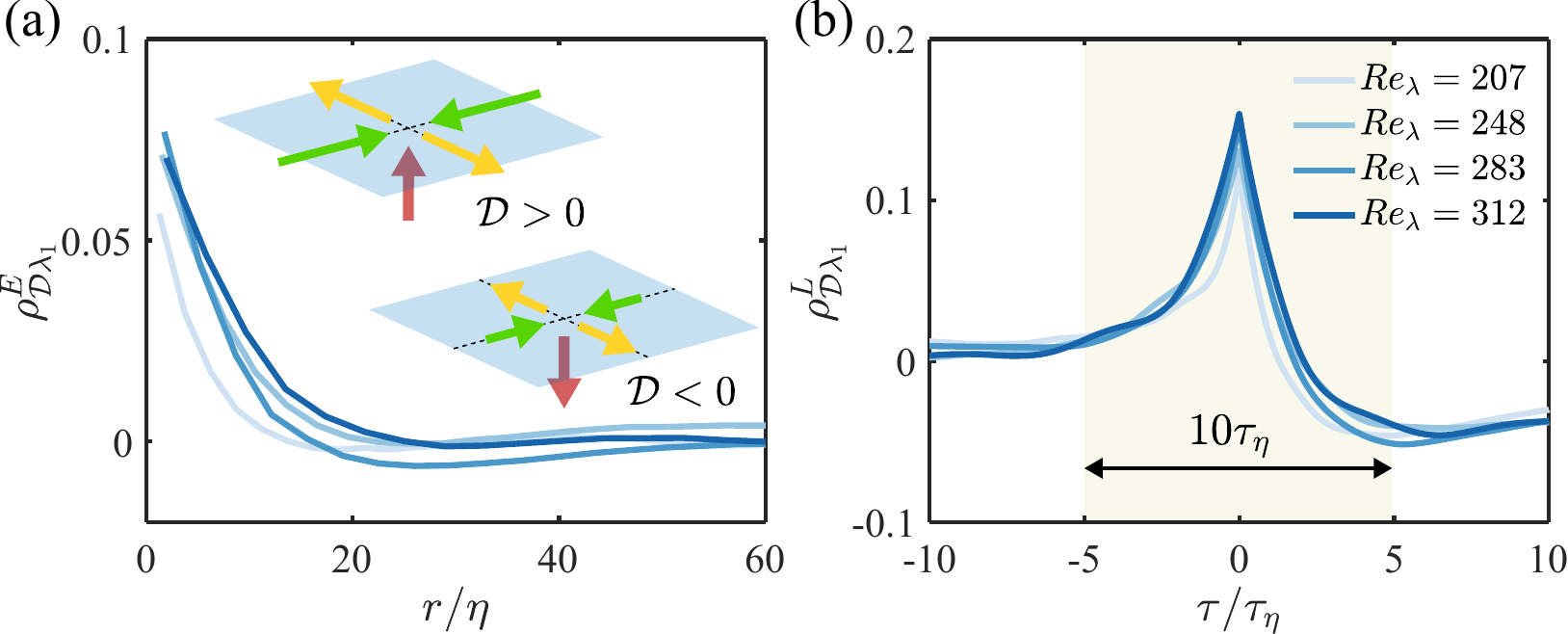}
    \caption{(a) The Eulerian cross-correlation function for the divergence and the larger eigenvalue of strain-rate tensor on the surface \(\lambda_{1}\) for different \(Re_{\lambda}\). This panel shares the same legend with panel (b). The inset illustrates the relation between the positive/negative divergence (red arrows) and the strength of stretching (yellow arrows) and compression (green arrows) on the free surface. The length of arrows marks the magnitude of stretching and compression. (b) The Lagrangian cross-correlation for the divergence and \(\lambda_{1}\) for different \(Re_{\lambda}\). The green shaded area which cover a time scale of \(10\tau_{\eta}\) marks the time range when \(\lambda_{1}\) increases and decreases.}
    \label{fig_div_strain_cross}
\end{figure}

As shown in section \ref{sec_topology}, the surface divergence is associated to vorticity as well as to strain-rate, and in fact somewhat more significantly to the latter. To quantify this aspect, we consider the Eulerian cross-correlation between the surface divergence and \(\lambda_{1}\):
\begin{equation}
    \rho_{\mathcal{D}\lambda_{1}}^{E} = \frac{\langle\mathcal{D}'\left( \boldsymbol{x} \right)\lambda_{1}'\left( \boldsymbol{x} + r{\hat{\boldsymbol{e}}}_{r} \right)\rangle}{\mathcal{\langle D}'^{2}\rangle^{1/2}\langle\lambda_{1}'^{2}\rangle^{1/2} }.
\end{equation}
Figure \ref{fig_div_strain_cross}(a) plots this quantity for all cases, showing a positive correlation. This indicates that a larger positive surface divergence (stronger source) is likely to be associated with large (i.e., above average) strain-rate along the surface stretching direction. As \(\lambda_{1}\) and \(\lambda_{2}\) are highly anti-correlated (see Figure \ref{fig_strain_eigen_pdf}(b)), a compression of comparable magnitude is also likely to occur in the surface-parallel direction perpendicular to the one of stretching (with the compression slightly weaker than the stretching to satisfy incompressibility). This flow pattern is illustrated by the upper schematic in Figure \ref{fig_div_strain_cross}(a). On the other hand, a negative divergence is likely associated with weak (i.e., below average) strain-rate along the surface accompanied by a weak compression perpendicular to it, as also sketched in the lower schematic of Figure \ref{fig_div_strain_cross}(a). The cross-correlation has a characteristic scale comparable to the length scale of divergence, similar as \(\rho_{\mathcal{D}|\omega|}^{E}\) (Figure \ref{fig_div_vor_cross}(a)), though the magnitude of \(\rho_{\mathcal{D}\lambda_{1}}^{E}\) is somewhat larger. This is consistent with the observation that divergent regions overlap slightly more with high-strain-rate regions compared to high-vorticity regions (Figure \ref{fig_topology}(d)).

Finally, we examine the Lagrangian cross-correlation between divergence and \(\lambda_{1}\) (Figure \ref{fig_div_strain_cross}(b))
\begin{equation}
    \rho_{\mathcal{D}\lambda_{1}}^{L} = \frac{\langle\mathcal{D}'(t)\lambda_{1}'(t + \tau)\rangle}{\mathcal{\langle D}'^{2}\rangle^{1/2}\langle\lambda_{1}'^{2}\rangle^{1/2}}.
\end{equation}
Before a strong divergence event occurs (i.e., \(\tau < 0\)), \(\mathcal{D}\) and \(\lambda_{1}\) grow together as indicated by the increasing correlation \(\rho_{\mathcal{D}\lambda_{1}}^{L}\). Considering the strong anti-correlation between \(\lambda_{1}\) and \(\lambda_{2}\), this implies that the formation of a source is preceded by an increase in magnitude of the stretching-compression saddle along the surface. After \(\lambda_{1}\) reaches its maximum at \(\tau = 0\), it decreases and slowly approaches its mean with the stretching-compression saddle recovering to its average magnitude. For all \(Re_{\lambda}\), the duration of significant correlation is about \(10\tau_{\eta}\) (green shaded area), which is consistent with the lifetime of high-strain-rate events (Figure \ref{fig_lag_autocorr}(c)).

\subsection{Clustering on the free surface}\label{sec_cluster}

We finally examine the clustering of the floating particles along the surface. This process, originating from the compressible nature of the free surface, differs from the clustering of inertial particles in incompressible turbulence \citep{balachandar2010turbulent,brandt2022particle}. The latter results from the fact that inertial particles depart from the pathlines of fluid parcels whose fluctuations they cannot follow, thus leading to a compressible field \citep{maxey1987gravitational}. On the other hand, floating particles cannot follow the surface pathlines entering the bulk.

\begin{figure}
    \centering
    \includegraphics[width=0.95\linewidth]{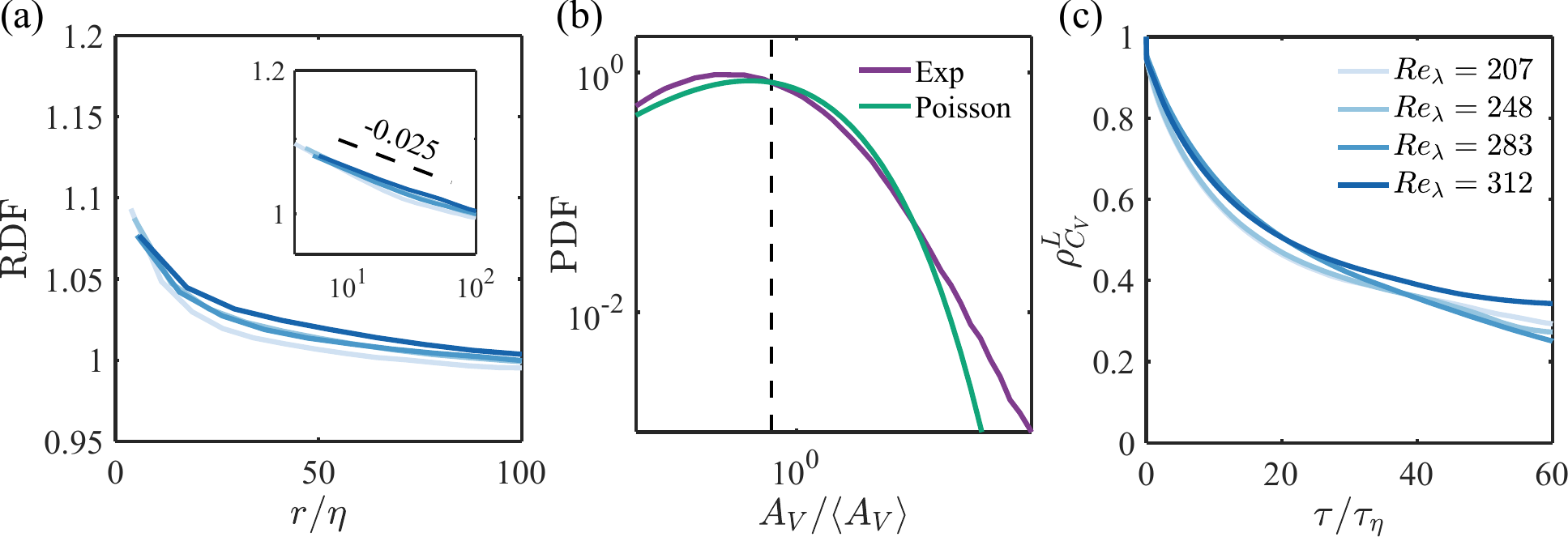}
    \caption{(a) The RDF of floating particles on the free surface for various \(Re_{\lambda}\). The inset shows the same figure with the horizontal and vertical axes in logarithmic scales. This panel shares the same legend with panel (c). (b) The PDF of the area of Voronoi cells around floating particle for the case \(Re_{\lambda} = 312\). The purple line shows the PDF for a random Poisson process. The black dashed line at \(A_{V}/\left\langle A_{V} \right\rangle = 0.83\) marks the crossing point between both curves. (c) The Lagrangian autocorrelation of the partile concentration for various \(Re_{\lambda}\).}
    \label{fig_clustering}
\end{figure}

Figure \ref{fig_clustering}(a) shows the radial distribution function (RDF) of floating particles defined as \(g(r) = \left( N_{r}/A_{r} \right)/ \left( N_{t}/A_{FOV} \right)\), where \(N_{r}\) is the number of particles within a narrow circular ring with radius \(r\) and area \(A_{r}\), and \(N_{t}\) is the total number of particles in the FOV. The RDF quantifies the local concentration around a generic particle, and thus values larger than unity indicate the formation of clusters over a certain length scale. In the present case, the values indicate moderate degree of clustering. This is compatible with the weak compressibility we observe. An exponential fit to the data yields a characteristic length scale of the clustering around \(30\eta\), close to the length scale \(\sim 20\eta\) found in the simulations by \cite{schumacher2002clustering}. The fact that the cluster length scale is significantly larger than that of divergence (Figure \ref{fig_euler_autocorr}(a)) suggests the clustering is also affected by the large-scale motion on the free surface. The inset of Figure \ref{fig_clustering}(a) displays the same RDF in logarithmic scale. While the range is not sufficient for a conclusive statement, the data is compatible with a power-law decay, which would indicate spatial self-similarity of the concentration field. This would be in turn consistent with previous works that show how floating particles cluster over fractal sets \citep{boffetta2006multifractal,larkin2009power}.

The clustering of floating particles is also characterized using the Voronoi tessellation method \citep{monchaux2010preferential}. Figure \ref{fig_clustering}(b) displays the PDF of the area of the Voronoi cells \(A_{V}\) for \(Re_{\lambda} = 312\), compared with the PDF that approximates the distribution of scattered particles in a random Poisson process \citep{ferenc2007size}. We note the data only covers a limited range down to \(A_{V}/\left\langle A_{V} \right\rangle \approx 0.3\). This corresponds to \(5 \times 5\) pixels in raw images, slightly larger than the particle size. Voronoi cells smaller than this criterion tend to have larger uncertainty. Nevertheless, the area PDF clearly shows higher probability at \(A_{V}/\left\langle A_{V} \right\rangle < 0.83\) and \(A_{V}/\left\langle A_{V} \right\rangle > 2\), indicating the occurrence of clusters and voids, respectively. The degree of clustering is quantified by calculating the standard deviation of the PDF, \(\sigma_{AV}\), and comparing with that of the random Poisson process, \(\sigma_{rpp} \sim 0.53\). The ratio \(\sigma_{AV}/\sigma_{rpp} \sim 1.6\) confirms the moderate intensity of the clustering.

The time scale of the clustering is further characterized by calculating its Lagrangian autocorrelation function
\begin{equation}
    \rho_{C_{V}}^{L} = \frac{\langle C_{V}'(t)C_{V}'(t + \tau)\rangle}{\langle C_{V}'^{2}\rangle},
\end{equation}
as shown in Figure \ref{fig_clustering}(c), where \(C_{V} = 1/A_{V}\) denotes the local concentration of particles. Results for different \(Re_{\lambda}\) collapse and exhibit a similar time scale around \(\sim 40\tau_{\eta}\). This is significantly larger than the characteristic time scale of the divergence and is close to the integral time scale, highlighting again the role of the large-scale motions. This result is consistent with the observation by \cite{lovecchio2013time} who reported clusters evolving over a time scale similar to and even larger than the integral time scale of the underlying turbulence.

It is worth mentioning that although the formation of clusters is associated with the compressible surface velocity field, the distribution of clusters is not expected to exhibit strong connection with the instantaneous divergence field. Instead, clusters (voids) emerge where persistent sinks (sources) are present, i.e., the time history of the surface divergence needs to be considered. This mechanism potentially elucidates the distinct temporal and spatial scales exhibited by the clusters compared with the divergence.

\section{Conclusions}\label{sec_conclusion}

In this work, we experimentally studied the small-scale dynamics of the free surface above homogeneous and isotropic turbulent water. We focus on a regime of negligible surface deformation. The experiment is conducted in a zero-mean flow turbulent water tank in which the turbulence is forced by two opposing placed jet arrays. A Taylor Reynolds number \(Re_{\lambda}\)= 207--312 is achieved and high-speed/high-resolution imaging is used to measure the free-surface flow. By seeding high-concentration floating particles, the surface velocity field is obtained by PTV and the velocity gradient tensor is calculated along each trajectory by a local least-square approach.

We first derive a kinematic relation for the energy dissipation rate on the free surface. By applying the free-surface boundary conditions and assuming small-scale homogeneity and isotropy of the flow, the dissipation rate can be written as a function of surface divergence and vorticity, highlighting its connection with the non-solenoidal nature of the surface and surface-normal vorticity.

The PDFs of divergence, vorticity and strain-rate collapse once normalized by the Kolmogorov scales over the considered range of \(Re_{\lambda}\). The symmetry of the PDF of divergence indicates that sources and sinks have similar strength. The two eigenvalues of the strain-rate tensor show clear anti-symmetry and anti-correlation, suggesting the stretching and compression along the free surface are equally likely and intense, in contrast with 3D turbulence case in which stretching is predominant. The magnitude of these quantities is examined by plotting PDFs of the squares of them, in which power-law tails at small magnitude are evident. We show that this is due to the Gaussian core of the velocity gradient PDF. As a result, these squared quantities follow chi-square distribution of different orders based on their definitions.

The intense-divergence, intense-vorticity and intense-strain-rate structures are identified by a percolation technique. The PDFs of the area of the structures show power-law scaling (though over a limited size range) suggesting that these structures are self-similar. The intense-divergence structures have smaller area, whereas the intense-vorticity and intense-strain structures are more elongated. These observations are further confirmed by their instantaneous fields. Moreover, strong overlap between the intense-vorticity and intense-strain region is also observed.

To examine the energy at different scales, the second-order structure function along the free surface is considered. A clear \(r^{2/3}\) scaling is evident, consistent with the classic Kolmogorov theory. The scaling for the dissipation rate, on the other hand, is only preserved when the energy dissipation rate on the free surface is used, i.e., \(D_{LL} \sim \epsilon_{s}^{2/3}\). This leads to a factor \(C_{2s} \approx 3.5\) which deviates from the one for 3D turbulence. The plateau of \(D_{LL}\) at millimetric separation signals anomalously large relative velocities which are attributed to the compressible nature of the free-surface flow.

The scales of divergence, vorticity and strain are examined by calculating the Eulerian and Lagrangian autocorrelation functions. The results collapse after normalizing by Kolmogorov scales. The time scale and length scale of divergence are close to Kolmogorov scales, suggesting the divergence is driven by small-scale processes. On the other hand, the vorticity and strain-rate have larger length scales and are much longer-lived compared to divergence. This behavior emphasizes the difference between the free-surface turbulence and 3D turbulence, in which the time scales of vorticity and strain-rate are on the order of the Kolmogorov time scale.

The mutual correlation among the divergence, vorticity and strain-rate is explored by calculating the cross-correlation functions. Negative divergence events (sinks) are found to increase the magnitude of vorticity through a vortex-stretching process during which the surface-attached vortex is stretched by the downwelling. After this downwelling dissipates, the vortex diffuses and the vorticity decays. The evolution of the surface vorticity is characterized by the term \(\omega^{2}\mathcal{D}\), whose PDF is symmetric, in stark contrast with 3D turbulence where vortex stretching prevails. Moreover, upwelling events are likely to be associated with strong stretching/compression saddles along the free surface, while downwellings are associated with weak surface-parallel stretching/compression. The growth and decay of the saddle intensity during upwelling event is clearly illustrated by Lagrangian autocorrelations

Finally, the clustering of the floating particles due to surface divergence is examined. The RDF and the Voronoi tessellation method indicate moderate clustering, consistent with the weak compressibility we observe. The clusters exhibit characteristic spatial and temporal scales greater than those of the divergence, suggesting the former is directly affected by the large-scale motions. Taken together, the results of this study indicate that, in free-surface turbulence, the energetic scales leave a clearer imprint on the small-scales quantities compared to what usually observed in 3D turbulence.

We note that the surface contamination in the experiment could potentially affect the surface dynamics, including the surface-parallel fluctuation and surface divergence. However, this applies to other free-surface flows even when great care is taken in cleaning the surface. Our results are robust in that they do not change significantly with the time after the surface skimming. Still, it will be interesting to quantify differences in behavior with respect to situations in which the surface is completely devoid of surfactants. Such a study poses evident challenges for large-scale setups as the present one, and is beyond the scope of this work.
 
This work probes several fundamental aspects of free-surface flows, including the free-surface dissipation rate, the statistics and topology of velocity gradient, the Kolmogorov scaling in the inertial range, and the effect of divergence on surface properties. The results may further shed light on other associated physical processes. In particular, the enhanced intermittency of the velocity gradient is found to be associated with the nonlinear self-amplification \citep{qi2024restricted}; the balance of vortex stretching and compression might explain the direction of energy cascade \citep{pan1995numerical,lovecchio2015upscale,ruth2024structure}; and the new scaling in the second-order structure function might account for the distinct dispersion behavior of floating particles \citep{eckhardt2001turbulence,li2024relative}. Several questions that are outside the scope of this work may be better understood leveraging the present findings, such as the exchange of mass and energy between the surface and the bulk, the inter-scale energy flux, and the role of surface deformation on the dynamics of free-surface flow. Dedicated experiments that acquire data on the surface deformation, surface flow, and flow in the bulk are required to tackle such problems.



\backsection[Funding]{Funding from the Swiss National Science Foundation (project \# 200021-207318) is gratefully acknowledged.}

\backsection[Declaration of interests]{The authors report no conflict of interest.}

\backsection[Data availability statement]{All the data supporting this work are available from the corresponding author upon reasonable request.}

\backsection[Author ORCIDs]{Y. Qi, https://orcid.org/0009-0004-9858-9411; Y. Li, https://orcid.org/0000-0003-1318-7073; F. Coletti, https://orcid.org/0000-0001-5344-2476}


\bibliographystyle{jfm}
\bibliography{ref}

\begin{thebibliography}{85}
\expandafter\ifx\csname natexlab\endcsname\relax\def\natexlab#1{#1}\fi
\def\au#1{#1} \def\ed#1{#1} \def\yr#1{#1}\def\at#1{#1}\def\jt#1{\textit{#1}} \def\bt#1{#1}\def\bvol#1{\textbf{#1}} \def\vol#1{#1} \def\pg#1{#1} \def\publ#1{#1}\def\arxiv#1{#1}\def\org#1{#1}\def\st#1{\textit{#1}}

\bibitem[Babiker {\em et~al.\/}(2023)Babiker, Bjerkeb{\ae}k, Xuan, Shen \& Ellingsen]{babiker2023vortex}
{\sc \au{Babiker, Omer~M}, \au{Bjerkeb{\ae}k, Ivar}, \au{Xuan, Anqing}, \au{Shen, Lian} \& \au{Ellingsen, Simen~{\AA}}} \yr{2023}  \at{Vortex imprints on a free surface as proxy for surface divergence}.  \jt{Journal of Fluid Mechanics}  \bvol{964},  \pg{R2}.

\bibitem[Balachandar \& Eaton(2010)]{balachandar2010turbulent}
{\sc \au{Balachandar, S} \& \au{Eaton, John~K}} \yr{2010}  \at{Turbulent dispersed multiphase flow}.  \jt{Annual review of fluid mechanics}  \bvol{42}~(1),  \pg{111--133}.

\bibitem[Banerjee(1994)]{banerjee1994upwellings}
{\sc \au{Banerjee, Sanjoy}} \yr{1994}  \at{Upwellings, downdrafts, and whirlpools: Dominant structures in free surface turbulence} .

\bibitem[Bec {\em et~al.\/}(2010)Bec, Biferale, Cencini, Lanotte \& Toschi]{bec2010intermittency}
{\sc \au{Bec, J}, \au{Biferale, L}, \au{Cencini, M}, \au{Lanotte, AS} \& \au{Toschi, F}} \yr{2010}  \at{Intermittency in the velocity distribution of heavy particles in turbulence}.  \jt{Journal of Fluid Mechanics}  \bvol{646},  \pg{527--536}.

\bibitem[Bec {\em et~al.\/}(2024)Bec, Gustavsson \& Mehlig]{bec2024statistical}
{\sc \au{Bec, J}, \au{Gustavsson, K} \& \au{Mehlig, B}} \yr{2024}  \at{Statistical models for the dynamics of heavy particles in turbulence}.  \jt{Annual Review of Fluid Mechanics}  \bvol{56}~(1),  \pg{189--213}.

\bibitem[Bechlars \& Sandberg(2017)]{bechlars2017variation}
{\sc \au{Bechlars, P} \& \au{Sandberg, RD}} \yr{2017}  \at{Variation of enstrophy production and strain rotation relation in a turbulent boundary layer}.  \jt{Journal of Fluid Mechanics}  \bvol{812},  \pg{321--348}.

\bibitem[Berk \& Coletti(2021)]{berk2021dynamics}
{\sc \au{Berk, Tim} \& \au{Coletti, Filippo}} \yr{2021}  \at{Dynamics of small heavy particles in homogeneous turbulence: a lagrangian experimental study}.  \jt{Journal of Fluid Mechanics}  \bvol{917},  \pg{A47}.

\bibitem[Berk \& Coletti(2023)]{berk2023dynamics}
{\sc \au{Berk, Tim} \& \au{Coletti, Filippo}} \yr{2023}  \at{Dynamics and scaling of particle streaks in high-reynolds-number turbulent boundary layers}.  \jt{Journal of Fluid Mechanics}  \bvol{975},  \pg{A47}.

\bibitem[Betchov(1956)]{betchov1956inequality}
{\sc \au{Betchov, R}} \yr{1956}  \at{An inequality concerning the production of vorticity in isotropic turbulence}.  \jt{Journal of Fluid Mechanics}  \bvol{1}~(5),  \pg{497--504}.

\bibitem[Bewley {\em et~al.\/}(2013)Bewley, Saw \& Bodenschatz]{bewley2013observation}
{\sc \au{Bewley, Gregory~P}, \au{Saw, Ewe-Wei} \& \au{Bodenschatz, Eberhard}} \yr{2013}  \at{Observation of the sling effect}.  \jt{New Journal of Physics}  \bvol{15}~(8),  \pg{083051}.

\bibitem[Boffetta {\em et~al.\/}(2006)Boffetta, Davoudi \& De~Lillo]{boffetta2006multifractal}
{\sc \au{Boffetta, Guido}, \au{Davoudi, J} \& \au{De~Lillo, Filippo}} \yr{2006}  \at{Multifractal clustering of passive tracers on a surface flow}.  \jt{Europhysics Letters}  \bvol{74}~(1),  \pg{62}.

\bibitem[Boffetta {\em et~al.\/}(2004)Boffetta, De~Lillo \& Gamba]{boffetta2004large}
{\sc \au{Boffetta, Guido}, \au{De~Lillo, Filippo} \& \au{Gamba, A}} \yr{2004}  \at{Large scale inhomogeneity of inertial particles in turbulent flows}.  \jt{Physics of Fluids}  \bvol{16}~(4),  \pg{L20--L23}.

\bibitem[Brandt \& Coletti(2022)]{brandt2022particle}
{\sc \au{Brandt, Luca} \& \au{Coletti, Filippo}} \yr{2022}  \at{Particle-laden turbulence: progress and perspectives}.  \jt{Annual Review of Fluid Mechanics}  \bvol{54}~(1),  \pg{159--189}.

\bibitem[Brocchini \& Peregrine(2001)]{brocchini2001dynamics}
{\sc \au{Brocchini, Maurizio} \& \au{Peregrine, DH1871644}} \yr{2001}  \at{The dynamics of strong turbulence at free surfaces. part 1. description}.  \jt{Journal of Fluid Mechanics}  \bvol{449},  \pg{225--254}.

\bibitem[Brumley \& Jirka(1987)]{brumley1987near}
{\sc \au{Brumley, Blair~H} \& \au{Jirka, Gerhard~H}} \yr{1987}  \at{Near-surface turbulence in a grid-stirred tank}.  \jt{Journal of Fluid Mechanics}  \bvol{183},  \pg{235--263}.

\bibitem[Buxton \& Ganapathisubramani(2010)]{buxton2010amplification}
{\sc \au{Buxton, ORH} \& \au{Ganapathisubramani, B}} \yr{2010}  \at{Amplification of enstrophy in the far field of an axisymmetric turbulent jet}.  \jt{Journal of fluid mechanics}  \bvol{651},  \pg{483--502}.

\bibitem[Calmet \& Magnaudet(2003)]{calmet2003statistical}
{\sc \au{Calmet, Isabelle} \& \au{Magnaudet, Jacques}} \yr{2003}  \at{Statistical structure of high-reynolds-number turbulence close to the free surface of an open-channel flow}.  \jt{Journal of Fluid Mechanics}  \bvol{474},  \pg{355--378}.

\bibitem[Cardesa {\em et~al.\/}(2013)Cardesa, Mistry, Gan \& Dawson]{cardesa2013invariants}
{\sc \au{Cardesa, JI}, \au{Mistry, Dhiren}, \au{Gan, Lian} \& \au{Dawson, JR}} \yr{2013}  \at{Invariants of the reduced velocity gradient tensor in turbulent flows}.  \jt{Journal of Fluid Mechanics}  \bvol{716},  \pg{597--615}.

\bibitem[Carter {\em et~al.\/}(2016)Carter, Petersen, Amili \& Coletti]{carter2016generating}
{\sc \au{Carter, Douglas}, \au{Petersen, Alec}, \au{Amili, Omid} \& \au{Coletti, Filippo}} \yr{2016}  \at{Generating and controlling homogeneous air turbulence using random jet arrays}.  \jt{Experiments in Fluids}  \bvol{57},  \pg{1--15}.

\bibitem[Carter \& Coletti(2018)]{carter2018small}
{\sc \au{Carter, Douglas~W} \& \au{Coletti, Filippo}} \yr{2018}  \at{Small-scale structure and energy transfer in homogeneous turbulence}.  \jt{Journal of Fluid Mechanics}  \bvol{854},  \pg{505--543}.

\bibitem[Chickadel {\em et~al.\/}(2011)Chickadel, Talke, Horner-Devine \& Jessup]{chickadel2011infrared}
{\sc \au{Chickadel, C~Chris}, \au{Talke, Stefan~A}, \au{Horner-Devine, Alexander~R} \& \au{Jessup, Andrew~T}} \yr{2011}  \at{Infrared-based measurements of velocity, turbulent kinetic energy, and dissipation at the water surface in a tidal river}.  \jt{IEEE Geoscience and Remote Sensing Letters}  \bvol{8}~(5),  \pg{849--853}.

\bibitem[Cressman {\em et~al.\/}(2004)Cressman, Davoudi, Goldburg \& Schumacher]{cressman2004eulerian}
{\sc \au{Cressman, John~R}, \au{Davoudi, Jahanshah}, \au{Goldburg, Walter~I} \& \au{Schumacher, J{\"o}rg}} \yr{2004}  \at{Eulerian and lagrangian studies in surface flow turbulence}.  \jt{New Journal of Physics}  \bvol{6}~(1),  \pg{53}.

\bibitem[Davidson(2015)]{davidson2015turbulence}
{\sc \au{Davidson, Peter}} \yr{2015} {\em Turbulence: an introduction for scientists and engineers\/}.  \publ{Oxford University Press, USA}.

\bibitem[Deike(2022)]{deike2022mass}
{\sc \au{Deike, Luc}} \yr{2022}  \at{Mass transfer at the ocean--atmosphere interface: the role of wave breaking, droplets, and bubbles}.  \jt{Annual Review of Fluid Mechanics}  \bvol{54},  \pg{191--224}.

\bibitem[Dong {\em et~al.\/}(2017)Dong, Lozano-Dur{\'a}n, Sekimoto \& Jim{\'e}nez]{dong2017coherent}
{\sc \au{Dong, Siwei}, \au{Lozano-Dur{\'a}n, Adri{\'a}n}, \au{Sekimoto, Atsushi} \& \au{Jim{\'e}nez, Javier}} \yr{2017}  \at{Coherent structures in statistically stationary homogeneous shear turbulence}.  \jt{Journal of Fluid Mechanics}  \bvol{816},  \pg{167--208}.

\bibitem[Durham {\em et~al.\/}(2013)Durham, Climent, Barry, De~Lillo, Boffetta, Cencini \& Stocker]{durham2013turbulence}
{\sc \au{Durham, William~M}, \au{Climent, Eric}, \au{Barry, Michael}, \au{De~Lillo, Filippo}, \au{Boffetta, Guido}, \au{Cencini, Massimo} \& \au{Stocker, Roman}} \yr{2013}  \at{Turbulence drives microscale patches of motile phytoplankton}.  \jt{Nature communications}  \bvol{4}~(1),  \pg{2148}.

\bibitem[Eckhardt \& Schumacher(2001)]{eckhardt2001turbulence}
{\sc \au{Eckhardt, Bruno} \& \au{Schumacher, J{\"o}rg}} \yr{2001}  \at{Turbulence and passive scalar transport in a free-slip surface}.  \jt{Physical Review E}  \bvol{64}~(1),  \pg{016314}.

\bibitem[van Emmerik \& Schwarz(2020)]{van2020plastic}
{\sc \au{van Emmerik, Tim} \& \au{Schwarz, Anna}} \yr{2020}  \at{Plastic debris in rivers}.  \jt{Wiley Interdisciplinary Reviews: Water}  \bvol{7}~(1),  \pg{e1398}.

\bibitem[Esteban {\em et~al.\/}(2019)Esteban, Shrimpton \& Ganapathisubramani]{esteban2019laboratory}
{\sc \au{Esteban, L~Blay}, \au{Shrimpton, JS} \& \au{Ganapathisubramani, Bharathram}} \yr{2019}  \at{Laboratory experiments on the temporal decay of homogeneous anisotropic turbulence}.  \jt{Journal of Fluid Mechanics}  \bvol{862},  \pg{99--127}.

\bibitem[Ferenc \& N{\'e}da(2007)]{ferenc2007size}
{\sc \au{Ferenc, J{\'a}rai-Szab{\'o}} \& \au{N{\'e}da, Zolt{\'a}n}} \yr{2007}  \at{On the size distribution of poisson voronoi cells}.  \jt{Physica A: Statistical Mechanics and its Applications}  \bvol{385}~(2),  \pg{518--526}.

\bibitem[George \& Hussein(1991)]{george1991locally}
{\sc \au{George, William~K} \& \au{Hussein, Hussein~J}} \yr{1991}  \at{Locally axisymmetric turbulence}.  \jt{Journal of Fluid Mechanics}  \bvol{233},  \pg{1--23}.

\bibitem[Goldburg {\em et~al.\/}(2001)Goldburg, Cressman, V{\"o}r{\"o}s, Eckhardt \& Schumacher]{goldburg2001turbulence}
{\sc \au{Goldburg, WI}, \au{Cressman, JR}, \au{V{\"o}r{\"o}s, Z}, \au{Eckhardt, B} \& \au{Schumacher, J}} \yr{2001}  \at{Turbulence in a free surface}.  \jt{Physical Review E}  \bvol{63}~(6),  \pg{065303}.

\bibitem[Guo \& Shen(2010)]{guo2010interaction}
{\sc \au{Guo, Xin} \& \au{Shen, Lian}} \yr{2010}  \at{Interaction of a deformable free surface with statistically steady homogeneous turbulence}.  \jt{Journal of fluid mechanics}  \bvol{658},  \pg{33--62}.

\bibitem[Herlina \& Jirka(2008)]{herlina2008experiments}
{\sc \au{Herlina} \& \au{Jirka, G.~H.}} \yr{2008}  \at{Experiments on gas transfer at the air--water interface induced by oscillating grid turbulence}.  \jt{Journal of Fluid Mechanics}  \bvol{594},  \pg{183--208}.

\bibitem[Herlina \& Wissink(2014)]{herlina2014direct}
{\sc \au{Herlina, H} \& \au{Wissink, JG}} \yr{2014}  \at{Direct numerical simulation of turbulent scalar transport across a flat surface}.  \jt{Journal of fluid mechanics}  \bvol{744},  \pg{217--249}.

\bibitem[Herlina \& Wissink(2019)]{herlina2019simulation}
{\sc \au{Herlina, H} \& \au{Wissink, JG}} \yr{2019}  \at{Simulation of air--water interfacial mass transfer driven by high-intensity isotropic turbulence}.  \jt{Journal of Fluid Mechanics}  \bvol{860},  \pg{419--440}.

\bibitem[Hunt \& Graham(1978)]{hunt1978free}
{\sc \au{Hunt, JCR} \& \au{Graham, JMR}} \yr{1978}  \at{Free-stream turbulence near plane boundaries}.  \jt{Journal of Fluid Mechanics}  \bvol{84}~(2),  \pg{209--235}.

\bibitem[J{\"a}hne \& Hau{\ss}ecker(1998)]{jahne1998air}
{\sc \au{J{\"a}hne, Bernd} \& \au{Hau{\ss}ecker, H}} \yr{1998}  \at{Air-water gas exchange}.  \jt{Annual Review of Fluid Mechanics}  \bvol{30}~(1),  \pg{443--468}.

\bibitem[Johnson \& Wilczek(2024)]{johnson2024multiscale}
{\sc \au{Johnson, Perry~L} \& \au{Wilczek, Michael}} \yr{2024}  \at{Multiscale velocity gradients in turbulence}.  \jt{Annual Review of Fluid Mechanics}  \bvol{56}~(1),  \pg{463--490}.

\bibitem[Kolmogorov(1941)]{kolmogorov1941local}
{\sc \au{Kolmogorov, Andrey~Nikolaevich}} \yr{1941}  \at{The local structure of turbulence in incompressible viscous fluid for very large reynolds numbers}.  \jt{Cr Acad. Sci. URSS}  \bvol{30},  \pg{301--305}.

\bibitem[Komori {\em et~al.\/}(1989)Komori, Murakami \& Ueda]{komori1989detection}
{\sc \au{Komori, Satoru}, \au{Murakami, Yasuhiro} \& \au{Ueda, Hiromasa}} \yr{1989}  \at{Detection of coherent structures associated with bursting events in an open-channel flow by a two-point measuring technique using two laser-doppler velocimeters}.  \jt{Physics of Fluids A: Fluid Dynamics}  \bvol{1}~(2),  \pg{339--348}.

\bibitem[Kumar {\em et~al.\/}(1998)Kumar, Gupta \& Banerjee]{kumar1998experimental}
{\sc \au{Kumar, S}, \au{Gupta, R} \& \au{Banerjee, S}} \yr{1998}  \at{An experimental investigation of the characteristics of free-surface turbulence in channel flow}.  \jt{Physics of fluids}  \bvol{10}~(2),  \pg{437--456}.

\bibitem[Larkin {\em et~al.\/}(2009)Larkin, Bandi, Pumir \& Goldburg]{larkin2009power}
{\sc \au{Larkin, Jason}, \au{Bandi, MM}, \au{Pumir, Alain} \& \au{Goldburg, Walter~I}} \yr{2009}  \at{Power-law distributions of particle concentration in free-surface flows}.  \jt{Physical Review E—Statistical, Nonlinear, and Soft Matter Physics}  \bvol{80}~(6),  \pg{066301}.

\bibitem[Li {\em et~al.\/}(2024)Li, Wang, Qi \& Coletti]{li2024relative}
{\sc \au{Li, Yaxing}, \au{Wang, Yifan}, \au{Qi, Yinghe} \& \au{Coletti, Filippo}} \yr{2024}  \at{Relative dispersion in free-surface turbulence}.  \jt{Journal of Fluid Mechanics, in press} .

\bibitem[Lindemann {\em et~al.\/}(2017)Lindemann, Visser \& Mariani]{lindemann2017dynamics}
{\sc \au{Lindemann, Christian}, \au{Visser, Andre} \& \au{Mariani, Patrizio}} \yr{2017}  \at{Dynamics of phytoplankton blooms in turbulent vortex cells}.  \jt{Journal of The Royal Society Interface}  \bvol{14}~(136),  \pg{20170453}.

\bibitem[Lovecchio {\em et~al.\/}(2013)Lovecchio, Marchioli \& Soldati]{lovecchio2013time}
{\sc \au{Lovecchio, Salvatore}, \au{Marchioli, Cristian} \& \au{Soldati, Alfredo}} \yr{2013}  \at{Time persistence of floating-particle clusters in free-surface turbulence}.  \jt{Physical Review E—Statistical, Nonlinear, and Soft Matter Physics}  \bvol{88}~(3),  \pg{033003}.

\bibitem[Lovecchio {\em et~al.\/}(2015)Lovecchio, Zonta \& Soldati]{lovecchio2015upscale}
{\sc \au{Lovecchio, Salvatore}, \au{Zonta, Francesco} \& \au{Soldati, Alfredo}} \yr{2015}  \at{Upscale energy transfer and flow topology in free-surface turbulence}.  \jt{Physical Review E}  \bvol{91}~(3),  \pg{033010}.

\bibitem[Lozano-Dur{\'a}n {\em et~al.\/}(2012)Lozano-Dur{\'a}n, Flores \& Jim{\'e}nez]{lozano2012three}
{\sc \au{Lozano-Dur{\'a}n, Adri{\'a}n}, \au{Flores, Oscar} \& \au{Jim{\'e}nez, Javier}} \yr{2012}  \at{The three-dimensional structure of momentum transfer in turbulent channels}.  \jt{Journal of Fluid Mechanics}  \bvol{694},  \pg{100--130}.

\bibitem[Magnaudet(2003)]{magnaudet2003high}
{\sc \au{Magnaudet, Jacques}} \yr{2003}  \at{High-reynolds-number turbulence in a shear-free boundary layer: revisiting the hunt--graham theory}.  \jt{Journal of Fluid Mechanics}  \bvol{484},  \pg{167--196}.

\bibitem[Maxey(1987)]{maxey1987gravitational}
{\sc \au{Maxey, Martin~R}} \yr{1987}  \at{The gravitational settling of aerosol particles in homogeneous turbulence and random flow fields}.  \jt{Journal of fluid mechanics}  \bvol{174},  \pg{441--465}.

\bibitem[McKenna \& McGillis(2004)]{mckenna2004role}
{\sc \au{McKenna, SP} \& \au{McGillis, WR}} \yr{2004}  \at{The role of free-surface turbulence and surfactants in air--water gas transfer}.  \jt{International Journal of Heat and Mass Transfer}  \bvol{47}~(3),  \pg{539--553}.

\bibitem[Meneveau(2011)]{meneveau2011lagrangian}
{\sc \au{Meneveau, Charles}} \yr{2011}  \at{Lagrangian dynamics and models of the velocity gradient tensor in turbulent flows}.  \jt{Annual Review of Fluid Mechanics}  \bvol{43}~(1),  \pg{219--245}.

\bibitem[Moisy \& Jim{\'e}nez(2004)]{moisy2004geometry}
{\sc \au{Moisy, Fr{\'e}d{\'e}ric} \& \au{Jim{\'e}nez, Javier}} \yr{2004}  \at{Geometry and clustering of intense structures in isotropic turbulence}.  \jt{Journal of fluid mechanics}  \bvol{513},  \pg{111--133}.

\bibitem[Monchaux {\em et~al.\/}(2010)Monchaux, Bourgoin \& Cartellier]{monchaux2010preferential}
{\sc \au{Monchaux, Romain}, \au{Bourgoin, Micka{\"e}l} \& \au{Cartellier, Alain}} \yr{2010}  \at{Preferential concentration of heavy particles: a vorono{\"\i} analysis}.  \jt{Physics of Fluids}  \bvol{22}~(10).

\bibitem[Mordant {\em et~al.\/}(2004)Mordant, Crawford \& Bodenschatz]{mordant2004experimental}
{\sc \au{Mordant, Nicolas}, \au{Crawford, Alice~M} \& \au{Bodenschatz, Eberhard}} \yr{2004}  \at{Experimental lagrangian acceleration probability density function measurement}.  \jt{Physica D: Nonlinear Phenomena}  \bvol{193}~(1-4),  \pg{245--251}.

\bibitem[Mountford \& Morales~Maqueda(2019)]{mountford2019eulerian}
{\sc \au{Mountford, Alethea~Sara} \& \au{Morales~Maqueda, MA}} \yr{2019}  \at{Eulerian modeling of the three-dimensional distribution of seven popular microplastic types in the global ocean}.  \jt{Journal of Geophysical Research: Oceans}  \bvol{124}~(12),  \pg{8558--8573}.

\bibitem[Mullin \& Dahm(2006)]{mullin2006dual}
{\sc \au{Mullin, John~A} \& \au{Dahm, Werner~JA}} \yr{2006}  \at{Dual-plane stereo particle image velocimetry measurements of velocity gradient tensor fields in turbulent shear flow. ii. experimental results}.  \jt{Physics of Fluids}  \bvol{18}~(3).

\bibitem[Nagaosa(1999)]{nagaosa1999direct}
{\sc \au{Nagaosa, Ryuichi}} \yr{1999}  \at{Direct numerical simulation of vortex structures and turbulent scalar transfer across a free surface in a fully developed turbulence}.  \jt{Physics of Fluids}  \bvol{11}~(6),  \pg{1581--1595}.

\bibitem[Nikora {\em et~al.\/}(2007)Nikora, Nokes, Veale, Davidson \& Jirka]{nikora2007large}
{\sc \au{Nikora, V}, \au{Nokes, R}, \au{Veale, W}, \au{Davidson, M} \& \au{Jirka, GH}} \yr{2007}  \at{Large-scale turbulent structure of uniform shallow free-surface flows}.  \jt{Environmental Fluid Mechanics}  \bvol{7},  \pg{159--172}.

\bibitem[Pan \& Banerjee(1995)]{pan1995numerical}
{\sc \au{Pan, Y} \& \au{Banerjee, S}} \yr{1995}  \at{A numerical study of free-surface turbulence in channel flow}.  \jt{Physics of Fluids}  \bvol{7}~(7),  \pg{1649--1664}.

\bibitem[Perot \& Moin(1995)]{perot1995shear}
{\sc \au{Perot, Blair} \& \au{Moin, Parviz}} \yr{1995}  \at{Shear-free turbulent boundary layers. part 1. physical insights into near-wall turbulence}.  \jt{Journal of Fluid Mechanics}  \bvol{295},  \pg{199--227}.

\bibitem[Petersen {\em et~al.\/}(2019)Petersen, Baker \& Coletti]{petersen2019experimental}
{\sc \au{Petersen, Alec~J}, \au{Baker, Lucia} \& \au{Coletti, Filippo}} \yr{2019}  \at{Experimental study of inertial particles clustering and settling in homogeneous turbulence}.  \jt{Journal of Fluid Mechanics}  \bvol{864},  \pg{925--970}.

\bibitem[Pumir {\em et~al.\/}(2013)Pumir, Bodenschatz \& Xu]{pumir2013tetrahedron}
{\sc \au{Pumir, Alain}, \au{Bodenschatz, Eberhard} \& \au{Xu, Haitao}} \yr{2013}  \at{Tetrahedron deformation and alignment of perceived vorticity and strain in a turbulent flow}.  \jt{Physics of Fluids}  \bvol{25}~(3),  \pg{035101}.

\bibitem[Qi {\em et~al.\/}(2022)Qi, Tan, Corbitt, Urbanik, Salibindla \& Ni]{qi2022fragmentation}
{\sc \au{Qi, Yinghe}, \au{Tan, Shiyong}, \au{Corbitt, Noah}, \au{Urbanik, Carl}, \au{Salibindla, Ashwanth~KR} \& \au{Ni, Rui}} \yr{2022}  \at{Fragmentation in turbulence by small eddies}.  \jt{Nature communications}  \bvol{13}~(1),  \pg{469}.

\bibitem[Qi {\em et~al.\/}(2024)Qi, Xu \& Coletti]{qi2024restricted}
{\sc \au{Qi, Yinghe}, \au{Xu, Zhenwei} \& \au{Coletti, Filippo}} \yr{2024}  \at{Restricted euler dynamics in free-surface turbulence}.  \jt{Under review} .

\bibitem[Richardson(1926)]{richardson1926atmospheric}
{\sc \au{Richardson, Lewis~Fry}} \yr{1926}  \at{Atmospheric diffusion shown on a distance-neighbour graph}.  \jt{Proceedings of the Royal Society of London. Series A, Containing Papers of a Mathematical and Physical Character}  \bvol{110}~(756),  \pg{709--737}.

\bibitem[Ruth \& Coletti(2024)]{ruth2024structure}
{\sc \au{Ruth, Daniel~J} \& \au{Coletti, Filippo}} \yr{2024}  \at{Structure and energy transfer in homogeneous turbulence below a free surface}.  \jt{arXiv preprint arXiv:2406.05889} .

\bibitem[Sanness~Salmon {\em et~al.\/}(2023)Sanness~Salmon, Baker, Kozarek \& Coletti]{sanness2023effect}
{\sc \au{Sanness~Salmon, Henri~R}, \au{Baker, Lucia~J}, \au{Kozarek, Jessica~L} \& \au{Coletti, Filippo}} \yr{2023}  \at{Effect of shape and size on the transport of floating particles on the free surface in a natural stream}.  \jt{Water Resources Research}  \bvol{59}~(10),  \pg{e2023WR035716}.

\bibitem[Schumacher \& Eckhardt(2002)]{schumacher2002clustering}
{\sc \au{Schumacher, J{\"o}rg} \& \au{Eckhardt, Bruno}} \yr{2002}  \at{Clustering dynamics of lagrangian tracers in free-surface flows}.  \jt{Physical Review E}  \bvol{66}~(1),  \pg{017303}.

\bibitem[Shen {\em et~al.\/}(1999)Shen, Zhang, Yue \& Triantafyllou]{shen1999surface}
{\sc \au{Shen, Lian}, \au{Zhang, Xiang}, \au{Yue, Dick~KP} \& \au{Triantafyllou, George~S}} \yr{1999}  \at{The surface layer for free-surface turbulent flows}.  \jt{Journal of Fluid Mechanics}  \bvol{386},  \pg{167--212}.

\bibitem[Sreenivasan(1991)]{sreenivasan1991fractals}
{\sc \au{Sreenivasan, KR1090334}} \yr{1991}  \at{Fractals and multifractals in fluid turbulence}.  \jt{Annual review of fluid mechanics}  \bvol{23}~(1),  \pg{539--604}.

\bibitem[Sreenivasan(1995)]{sreenivasan1995universality}
{\sc \au{Sreenivasan, Katepalli~R}} \yr{1995}  \at{On the universality of the kolmogorov constant}.  \jt{Physics of Fluids}  \bvol{7}~(11),  \pg{2778--2784}.

\bibitem[Sreenivasan \& Antonia(1997)]{sreenivasan1997phenomenology}
{\sc \au{Sreenivasan, Katepalli~R} \& \au{Antonia, Robert~A}} \yr{1997}  \at{The phenomenology of small-scale turbulence}.  \jt{Annual review of fluid mechanics}  \bvol{29}~(1),  \pg{435--472}.

\bibitem[Tamburrino \& Gulliver(2007)]{tamburrino2007free}
{\sc \au{Tamburrino, Aldo} \& \au{Gulliver, John~S}} \yr{2007}  \at{Free-surface visualization of streamwise vortices in a channel flow}.  \jt{Water Resources Research}  \bvol{43}~(11).

\bibitem[Teixeira \& Belcher(2000)]{teixeira2000dissipation}
{\sc \au{Teixeira, MAC} \& \au{Belcher, SE}} \yr{2000}  \at{Dissipation of shear-free turbulence near boundaries}.  \jt{Journal of Fluid Mechanics}  \bvol{422},  \pg{167--191}.

\bibitem[Thomas \& Hancock(1977)]{thomas1977grid}
{\sc \au{Thomas, NH} \& \au{Hancock, PE}} \yr{1977}  \at{Grid turbulence near a moving wall}.  \jt{Journal of Fluid Mechanics}  \bvol{82}~(3),  \pg{481--496}.

\bibitem[Turney \& Banerjee(2013)]{turney2013air}
{\sc \au{Turney, Damon~E} \& \au{Banerjee, Sanjoy}} \yr{2013}  \at{Air--water gas transfer and near-surface motions}.  \jt{Journal of Fluid Mechanics}  \bvol{733},  \pg{588--624}.

\bibitem[Uzkan \& Reynolds(1967)]{uzkan1967shear}
{\sc \au{Uzkan, T} \& \au{Reynolds, WC}} \yr{1967}  \at{A shear-free turbulent boundary layer}.  \jt{Journal of Fluid Mechanics}  \bvol{28}~(4),  \pg{803--821}.

\bibitem[Variano \& Cowen(2008)]{variano2008random}
{\sc \au{Variano, Evan~A} \& \au{Cowen, Edwin~A}} \yr{2008}  \at{A random-jet-stirred turbulence tank}.  \jt{Journal of Fluid Mechanics}  \bvol{604},  \pg{1--32}.

\bibitem[Variano \& Cowen(2013)]{variano2013turbulent}
{\sc \au{Variano, Evan~A} \& \au{Cowen, Edwin~A}} \yr{2013}  \at{Turbulent transport of a high-schmidt-number scalar near an air--water interface}.  \jt{Journal of Fluid Mechanics}  \bvol{731},  \pg{259--287}.

\bibitem[Veron(2015)]{veron2015ocean}
{\sc \au{Veron, Fabrice}} \yr{2015}  \at{Ocean spray}.  \jt{Annual Review of Fluid Mechanics}  \bvol{47},  \pg{507--538}.

\bibitem[Walker {\em et~al.\/}(1996)Walker, Leighton \& Garza-Rios]{walker1996shear}
{\sc \au{Walker, DT}, \au{Leighton, RI} \& \au{Garza-Rios, Luis~O}} \yr{1996}  \at{Shear-free turbulence near a flat free surface}.  \jt{Journal of fluid Mechanics}  \bvol{320},  \pg{19--51}.

\bibitem[Yeung {\em et~al.\/}(2012)Yeung, Donzis \& Sreenivasan]{yeung2012dissipation}
{\sc \au{Yeung, PK}, \au{Donzis, DA} \& \au{Sreenivasan, KR}} \yr{2012}  \at{Dissipation, enstrophy and pressure statistics in turbulence simulations at high reynolds numbers}.  \jt{Journal of Fluid Mechanics}  \bvol{700},  \pg{5--15}.

\bibitem[Yeung {\em et~al.\/}(2015)Yeung, Zhai \& Sreenivasan]{yeung2015extreme}
{\sc \au{Yeung, PK}, \au{Zhai, XM} \& \au{Sreenivasan, Katepalli~R}} \yr{2015}  \at{Extreme events in computational turbulence}.  \jt{Proceedings of the National Academy of Sciences}  \bvol{112}~(41),  \pg{12633--12638}.

\bibitem[Zhang(2017)]{zhang2017transport}
{\sc \au{Zhang, Hua}} \yr{2017}  \at{Transport of microplastics in coastal seas}.  \jt{Estuarine, Coastal and Shelf Science}  \bvol{199},  \pg{74--86}.

\end{thebibliography}

\end{document}